\documentclass[mnsc,nonblindrev]{informs3_hide}

\OneAndAHalfSpacedXI

\usepackage{enumitem}
\usepackage{bbding}
\usepackage{pifont}
\usepackage{booktabs}
\usepackage{subcaption}
\usepackage{hyperref}
\usepackage{fvextra}
\usepackage{csquotes}
\usepackage{tabularx}%
\usepackage{threeparttable}
\usepackage{siunitx} \usepackage{makecell}
\usepackage{multirow}
\usepackage{pdflscape}
\usepackage{afterpage}

\usepackage{natbib}
 \bibpunct[, ]{(}{)}{,}{a}{}{,}%
 \def\bibfont{\small}%

\EquationsNumberedThrough    %

\TheoremsNumberedThrough     %
\ECRepeatTheorems  %

\begin{document}

\RUNTITLE{Uncertainty-Adjusted Sorting for Asset Pricing with Machine Learning}

\TITLE{Uncertainty-Adjusted Sorting for Asset Pricing with Machine Learning}

\ARTICLEAUTHORS{%

\AUTHOR{Yan Liu}
\AFF{SEM, SIGS and SIEM, Tsinghua University, Shenzhen 518038, China, \EMAIL{liuyan@sem.tsinghua.edu.cn}}

\AUTHOR{Ye Luo}
\AFF{Faculty of Business and Economics, The University of Hong Kong, Pokfulam Road, Hong Kong SAR, \EMAIL{kurtluo@hku.edu}}

\AUTHOR{Zigan Wang}
\AFF{SEM, SIGS and SIEM, Tsinghua University, Shenzhen 518038, China, \EMAIL{wangzg@sem.tsinghua.edu.cn}}

\AUTHOR{Xiaowei Zhang}
\AFF{Department of Industrial Engineering and Decision Analytics, Hong Kong University of Science and Technology, Clear Water Bay, Hong Kong SAR, \EMAIL{xiaowei@ust.hk}}
} %

\ABSTRACT{Machine learning is central to empirical asset pricing, but portfolio construction still relies on point predictions and largely ignores asset-specific estimation uncertainty. We propose a simple change: sort assets using uncertainty-adjusted prediction bounds instead of point predictions alone. Across a broad set of ML models and a U.S. equity panel, this approach improves portfolio performance relative to point-prediction sorting. These gains persist even when bounds are built from partial or misspecified uncertainty information. They arise mainly from reduced volatility and are strongest for flexible machine learning models. Identification and robustness exercises show that these improvements are driven by asset-level rather than time or aggregate predictive uncertainty.
}%


\maketitle

\section{Introduction}
A large and rapidly expanding literature demonstrates that machine learning (ML) methods substantially improve out-of-sample asset return prediction relative to conventional linear benchmarks, and that these statistical gains often translate into economically meaningful portfolio performance. Seminal contributions such as \cite{GuKellyXiu20} document large Sharpe ratio improvements from nonlinear learners in U.S. equities, while subsequent work extends these findings to stochastic discount factor estimation \citep{ChenPelgerZhu24}, international equity markets \citep{LeippoldWangZhou22}, and bond return forecasting
\citep{KellyPruittSu19,BianchiBuchnerTamoni20}. Collectively, this literature establishes ML as a powerful tool for extracting conditional expected returns in environments characterized by noisy signals, nonlinear interactions, and pervasive multicollinearity.

\begin{figure}
    \begin{center}
    \caption{Point Predictions with Uncertainty-Adjusted Bounds at Alternative Quantile Levels: A Stock-Level Illustration} \label{fig:prediction-illus}
\includegraphics[width=0.9\textwidth]{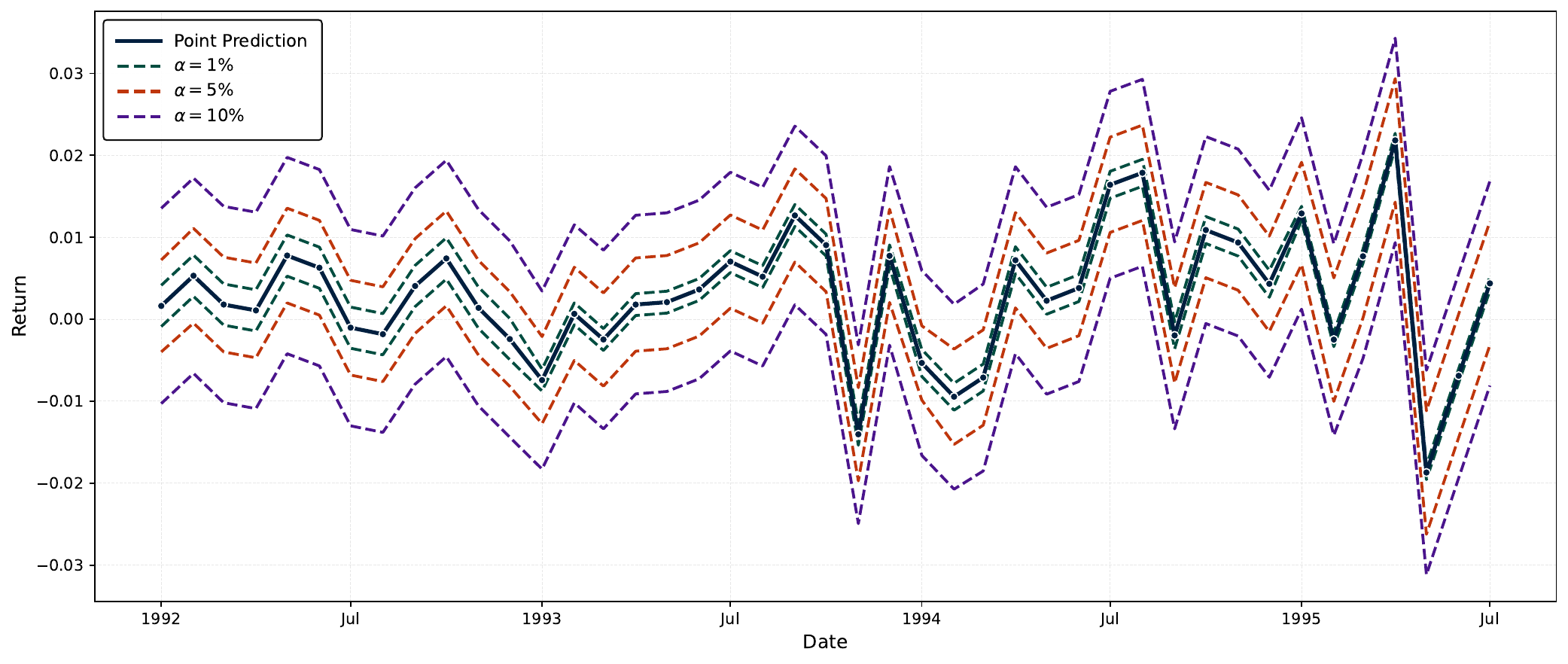}
    \end{center}
    \begin{tablenotes}
        \item \scriptsize{\emph{Note.} This figure plots point predictions of next-month excess returns together with uncertainty-adjusted upper and lower bounds for a randomly selected stock from the CRSP universe over the period January 1992 to July 1995. The solid line represents the model's point prediction. The dashed lines depict uncertainty-adjusted bounds constructed at quantile levels $\alpha$ = 1\%, 5\%, and 10\%, where a smaller $\alpha$ corresponds to narrower bounds. These bounds are obtained using rolling historical prediction residuals, without imposing parametric assumptions on the full residual distribution. The figure provides a time-series illustration of how predictive uncertainty varies over time and how alternative degrees of uncertainty adjustment affect the dispersion around point predictions. These uncertainty-adjusted bounds form the ranking signals used in the uncertainty-adjusted portfolio construction throughout the paper.}
    \end{tablenotes}
\end{figure}

Despite this progress, the dominant empirical practice in ML-based asset pricing remains centered on point predictions. Predicted returns are used to rank the cross-section and form long–short portfolios with little explicit assessment of asset-specific estimation uncertainty. This focus on point predictions stands in sharp contrast to modern statistical learning, where predictive uncertainty is treated as a first-class object \citep{EfronHastie16,VovkGammermanShafer22}.
Prediction intervals are routinely produced to quantify the dispersion of future outcomes, assess reliability, and guide decisions under uncertainty \citep{BatesAngelopoulosLeiMalikJordan21}. In asset pricing, this omission is especially consequential because portfolio construction is inherently a decision problem: investors allocate capital based not only on expected returns but also on the reliability of those expectations.\footnote{A recent paper by \cite{WangGaoHarveyLiuTao25} also considers how downstream decision-making could alter the first-stage prediction outcomes. Different from their paper, we focus on prediction accuracy, a primitive concept in statistical learning.} Figure~\ref{fig:prediction-illus} provides a stock-level illustration of predictive uncertainty. The figure plots point predictions together with alternative uncertainty-adjusted upper and lower bounds for a randomly selected stock, highlighting substantial time variation in predictive uncertainty even when point forecasts are similar. The construction of these bounds is described formally in Section~\ref{sec:interv-constr}.

This paper proposes and evaluates a simple yet powerful idea: augment ML point predictions with asset-specific prediction intervals and use these intervals---rather than point predictions---to sort assets for portfolio construction. We rank \emph{long positions by optimistic, uncertainty-adjusted predictions} and \emph{short positions by pessimistic, uncertainty-adjusted predictions}. Empirically, the resulting portfolios systematically outperform their point-prediction counterparts in terms of Sharpe ratios and volatility control.

Our contribution is distinct from two strands of the literature. First, unlike end-to-end or decision-focused learning frameworks that modify the training objective to internalize portfolio optimization directly \citep{ButlerKwon23,WangGaoHarveyLiuTao25}, we leave the prediction stage unchanged.
We use the same ML models trained under standard statistical objectives.
Second, unlike robust optimization approaches that require parametric assumptions on return distributions or covariance structure \citep{CostaIyengar23}, our method relies on distribution-free, residual-based predictive uncertainty that is directly estimated from the data.

The core innovation is not to develop a new ML model or to identify new predictors.
Instead, we change the sorting signal from a pure expectation to an uncertainty-adjusted sorting, catered to both traditional regression methods and more advanced ML methods, at partially internalizes uncertainty in a distribution-free manner. Crucially, the uncertainty-adjusted bounds employed in this paper serve as cross-sectional measures of estimation uncertainty, reflecting how reliably a given ML model predicts returns for each asset.

Using a U.S. equity panel spanning 1967--2016 and a comprehensive set of firm-level and macroeconomic predictors, we start from a standard panel prediction framework in which next-month returns are predicted using a rich set of firm characteristics and macroeconomic variables, estimated with a menu of ML models, including penalized linear models, dimension-reduction regressions, tree-based methods, and deep neural networks. For each asset $i$ and month $t$, we obtain a point prediction $\widehat{\mu}_{i,t}$ and construct uncertainty-adjusted bounds by calibrating historical out-of-sample residuals within a rolling window.
For a target coverage $\alpha$, the bounds form a prediction interval that is symmetric around the point prediction and takes the form:
\[
\text{PI}_{i,t}(\alpha) = \big[\, \widehat{\mu}_{i,t} - \widehat{q}_{i,t}(\alpha),\; \widehat{\mu}_{i,t} + \widehat{q}_{i,t}({\alpha}) \,\big],
\]
where $\widehat{q}_{i,t}(\alpha)$ is an estimate of  $\alpha$-quantile of (absolute) residuals constructed under a rolling window design.

Empirically, across a broad set of ML models and consistent with prior literature \citep{GuKellyXiu20}, we first show that the benchmark point-prediction sorting delivers strong performance in both long-short portfolios and index-relative long-only (excess-return) portfolios. For example, the Principal Components Regression (PCR) earns an annual return of 25.58\% and Sharpe ratio of 1.22, alongside Newey--West $t$-statistic of 5.45 \citep{NeweyWest87} and sizeable factor-adjusted alphas. Moreover, index-relative long-only excess returns are also substantial. We further show that these results remain economically meaningful after incorporating transaction costs: post-cost long–short returns and Sharpe ratios remain high for leading models, with PCR still delivering a Sharpe ratio of 0.97 after costs; post-cost long-only index-relative returns remain positive and statistically meaningful for most better-performing models.

We next move beyond the benchmark case and conduct analyses by incorporating the residual-quantile-based uncertainty-adjusted bounds as measures of predictive uncertainty. In each month $t$, we sort stocks by the upper bound $\widehat{\mu}_{i,t} + \widehat{q}_{i,t}({\alpha})$ to form the long leg and sort by the lower bound $\widehat{\mu}_{i,t} - \widehat{q}_{t}({\alpha})$ to form the short leg, using deciles in both cases. This differs from the benchmark point-prediction strategy, which sorts only by $\widehat{\mu}_{i,t}$.

We show that uncertainty-adjusted sorting can improve Sharpe ratios and $t$-statistics for most of ten models, where Sharpe ratios rise markedly when the uncertainty-adjusted sorting is used.
For example, the Sharpe ratio of the PCR increases to 1.56 under the 5\% quantile level. The performances of all five neural-network models have improved. For example, the Sharpe ratio of one neural network strategy (NN1) increases from 1.48 to 1.86.

We then study why the uncertainty-adjusted sorting helps. We define ``ranking improvements'' on the long and short sides as the difference between the uncertainty-adjusted rank of a stock and the point-prediction rank, and we regress these improvements on firm characteristics, macroeconomic conditions, and their interactions. The results indicate that ranking improvements are systematically related to firm size and value characteristics, and are strongly state-dependent with respect to macro/uncertainty conditions (e.g., volatility index).

Furthermore, we show that the performance gains from uncertainty-adjusted portfolios depend critically on correctly assigning predictive uncertainty to the assets and periods for which it is estimated. When this asset-level or time-series correspondence is disrupted, the gains are systematically attenuated—most sharply when uncertainty is misassigned across assets, and more moderately when it is misaligned over time. These results rule out purely mechanical explanations and underscore the economic importance of asset-specific predictive uncertainty.

Finally, we show that even partial information about predictive uncertainty is sufficient to improve portfolio performance. When uncertainty-adjusted bounds are constructed under a simple normality assumption using only the residual variance---without access to the full empirical residual distribution---the resulting portfolios still dominate point-prediction strategies in Sharpe ratio terms. This finding underscores the robustness and practical relevance of incorporating uncertainty information at the portfolio construction stage.

Our results speak to two literatures that have largely advanced in parallel. First, the ML asset pricing literature has established the predictive and economic value of nonlinear models but has largely treated predictions as point objects. Second, predictive inference literature, by contrast, has developed rigorous tools for uncertainty quantification but has rarely examined their implications for asset allocation. By integrating uncertainty-adjusted bounds into portfolio sorting, we provide a decision-relevant channel through which predictive uncertainty affects economic outcomes, without altering the underlying ML models or imposing strong distributional assumptions.

In sum, this paper shows that how predictions are used can matter as much as how they are generated. Incorporating even coarse measures of predictive uncertainty into portfolio sorting systematically improves performance, robustness, and interpretability in ML-based asset pricing.

\section{Literature Review}\label{sec:lit}

\subsection{Machine Learning in Empirical Asset Pricing}

Early studies of return predictability relied primarily on linear regressions using a small set of lagged predictors, such as dividend, price ratios, interest rates, and valuation measures. \cite{WelchGoyal08} provide a comprehensive reassessment and document that many prominent predictors perform poorly out-of-sample, highlighting instability and model uncertainty as central challenges in equity premium prediction.
These findings motivate the search for methods capable of handling high-dimensional predictors while mitigating overfitting and parameter instability.

A central development in empirical asset pricing over the past decade is the widespread adoption of ML tools to model nonlinearities, interactions, and high-dimensional predictor sets in return forecasting. Relative to traditional linear predictive regressions, modern ML methods are capable of flexibly approximating complex conditional mean functions and have been shown to deliver meaningful out-of-sample improvements in predictive accuracy. Importantly, this literature evaluates ML not only through statistical fit but also through its economic value, typically assessed via portfolio sorts and long--short trading strategies.

\cite{GuKellyXiu20} provide foundational evidence that ML models substantially improve the measurement of expected returns by capturing nonlinear interactions among firm characteristics. They document large economic gains in long--short portfolios, with particularly strong performance for tree-based methods and neural networks. Building on this insight, \cite{ChenPelgerZhu24} embed economic structure into deep learning by incorporating no-arbitrage restrictions into the estimation of stochastic discount factors, improving both interpretability and performance. \cite{AvramovChengMetzker23} further examine the robustness of ML signals under economic restrictions and transaction costs, emphasizing that profitability is closely linked to limits to arbitrage.

Taken together, these studies convincingly demonstrate that ML can play a central role in empirical asset pricing by extracting economically meaningful signals from high-dimensional data. At the same time, a common feature across this literature is a near-exclusive reliance on point predictions when constructing portfolios. Even when concerns about overfitting, instability, or trading frictions are explicitly addressed, predictive uncertainty remains implicit and is not incorporated at the asset level into the portfolio sorting rule itself.

A canonical empirical design in the ML asset pricing literature, adopted in influential studies \citep{GuKellyXiu20} and followed in this paper, proceeds in two steps.
First, a prediction function $\widehat{f}(x)$ is estimated to map predictors $x$---including firm characteristics and, in some cases, macroeconomic state variables---into one-period-ahead expected returns. Second, the resulting point predictions are used to rank stocks into deciles, forming long--short portfolios that go long the highest-ranked assets and short the lowest-ranked ones. This framework is powerful, transparent, and well suited for evaluating the economic relevance of predictive signals. However, it treats predicted returns as scalar sorting signals and leaves uncertainty in the predictions implicit.

Our study maintains the standard  evaluation framework for ML-based portfolios used in this literature but extends it in a new direction. Rather than modifying the learning objective or imposing additional economic constraints on the estimation stage, we augment the sorting signal by incorporating asset-specific predictive uncertainty.
By embedding uncertainty directly into the cross-sectional ranking that governs long and short positions, our approach complements existing ML asset pricing frameworks and addresses estimation uncertainty in portfolio construction, a dimension that has thus far received limited attention.

\subsection{Distribution-Free Predictive Inference}

A separate and rapidly growing literature develops distribution-free methods for predictive inference, focusing on the construction of uncertainty-adjusted bounds with finite-sample validity under minimal assumptions. The central objective of this literature is to construct a set-valued prediction rule $C(x)$ such that, for a new observation $(X_{n+1},Y_{n+1})$, $\Pr\big(Y_{n+1}\in C(X_{n+1})\big)\ge \alpha $
without relying on parametric assumptions about the data-generating process. This distribution-free guarantee distinguishes predictive inference from classical parametric uncertainty-adjusted bounds and makes it particularly attractive in high-dimensional and nonlinear settings.

Conformal prediction provides a canonical framework for achieving such guarantees. Originating in \cite{ShaferVovk08}, conformal methods construct prediction sets based on empirical ranks or quantiles of conformity scores---most commonly absolute residuals from a fitted model. \cite{AngelopoulosBarberBates25} provide a comprehensive theoretical treatment of conformal prediction, unifying residual-based, quantile-based, and cross-validation–based procedures within a permutation-inference framework with exact finite-sample guarantees. Under an exchangeability assumption, these procedures deliver exact finite-sample marginal coverage, regardless of the complexity of the underlying prediction algorithm. A large body of subsequent work extends the basic conformal framework along several dimensions. Cross-conformal prediction \citep{vovk2015cross} improves computational feasibility by aggregating conformal scores across data splits, while conformal predictive distributions \citep{vovk2018conformal} provide a distributional analogue that yields full predictive distributions rather than interval-valued sets. The jackknife+ method \citep{BarberCandesRamdasTibshirani21} further refines these ideas by using leave-one-out style predictions to obtain nonasymptotic coverage guarantees with improved finite-sample performance.

More recent research examines the conditional properties of distribution-free uncertainty-adjusted bounds. \cite{SteinbergerLeeb23}, the paper that we mainly follow for the quantile-of-residual methodology, study uncertainty-adjusted bounds constructed from cross-validation residuals and show that, for sufficiently stable learning algorithms, such intervals can achieve approximate conditional validity even in high-dimensional environments. This line of work highlights the role of algorithmic stability and cross-validation schemes in bridging the gap between marginal and conditional coverage guarantees.

An important practical consideration is that many prediction-inference methods rely on exchangeability \citep{BarberCandesRamdasTibshirani23}, an assumption that does not literally hold in time-series environments such as asset returns.
Modern approaches address this limitation by adapting conformal and residual-based methods to rolling, sequential, or block-based schemes, thereby preserving approximate validity while accommodating temporal dependence \citep{ChernozhukovWuthrichZhu21,XuXie23Conformal,XuXie23Sequential,OliveiraOrensteinRamosRomano24}.

Despite their methodological sophistication, these distribution-free inference techniques are primarily designed for statistical inference and coverage control, rather than for economic decision-making. In particular, uncertainty-adjusted bounds are typically reported as diagnostics of predictive uncertainty, not as objects that directly influence portfolio choice or cross-sectional ranking. As a result, while the predictive inference literature provides powerful tools for quantifying uncertainty, it leaves open the question of how such uncertainty should be incorporated into portfolio construction and asset pricing applications.

\subsection{The Gap and Our Contribution}

Taken together, the ML asset pricing literature and the predictive inference literature have made substantial but largely parallel advances. On the one hand, ML methods have been shown to significantly improve the measurement of expected returns by capturing nonlinearities and high-dimensional interactions among firm characteristics, with economically meaningful gains documented in long--short portfolios and uncertainty-adjusted performance metrics. On the other hand, the predictive inference literature has developed powerful distribution-free tools---such as conformal prediction, cross-conformal methods, and the jackknife+---to quantify predictive uncertainty and construct uncertainty-adjusted bounds with finite-sample validity under minimal assumptions.

Despite these advances, there is limited work that explicitly combines state-of-the-art ML return prediction with uncertainty quantification in a way that alters the portfolio decision rule itself. In existing ML asset pricing studies, predictive uncertainty is typically left implicit: predictions are treated as scalar sorting signals, and uncertainty-adjusted bounds---when constructed---are reported as diagnostics rather than as objects that influence portfolio formation. Conversely, the predictive inference literature focuses primarily on statistical coverage and calibration properties of uncertainty-adjusted bounds, with little attention to how such uncertainty-adjusted bounds should be incorporated into economic decision-making, particularly in cross-sectional portfolio selection problems.

This paper fills this gap by bridging these two strands in a novel and economically meaningful way. Rather than aiming to construct uncertainty-adjusted bounds with formal coverage guarantees or evaluating their inferential properties, we repurpose uncertainty information as an economic object---a cross-sectional, asset-specific measure of estimation uncertainty that directly enters portfolio construction. Specifically, we show that uncertainty-adjusted bounds can be used to construct upper- and lower-bound sorting signals, which replace point predictions in the sorting rule that determines long and short positions.

By integrating uncertainty-adjusted bounds into the ranking stage, we provide a simple, model-agnostic mechanism to discipline ML predictions. This approach preserves the strengths of ML in estimating conditional mean returns, while explicitly accounting for heterogeneity in predictive reliability across assets. The resulting portfolio strategies exhibit improved uncertainty-adjusted performance and display systematic heterogeneity linked to firm characteristics and macroeconomic states, shedding light on when and why uncertainty-adjusted sorting is most valuable.

The key novelty of our contribution is therefore economic rather than inferential. We do not treat uncertainty-adjusted bounds as an afterthought or as a purely statistical diagnostic. Instead, we embed them directly into the cross-sectional sorting mechanism that maps predictions into realized trading strategies. In doing so, our framework complements existing work that emphasizes economic restrictions, limits to arbitrage, and transaction costs, and offers a new perspective on the sources of performance gains in ML--based asset pricing strategies.

\section{Methodology}\label{sec:meth}

\subsection{Panel Prediction Framework and Information Set}

We adopt a unified panel-data prediction framework:
\[
r_{i,t+1} = f(x_{i,t}) + \varepsilon_{i,t+1},
\]
where $r_{i,t+1}$ is the one-month-ahead return of stock $i$, $x_{i,t}$ is a vector of contemporaneous firm characteristics and potentially macroeconomic state variables, and $f(\cdot)$ is an unknown prediction function estimated by different ML methods, and $\epsilon_{i,t+1}$ is the unobserved residual that are assumed to be independent across time.

We impose two modeling restrictions to align the estimation problem with standard empirical asset-pricing practice.
First, cross-sectional and temporal homogeneity: $f(\cdot)$ does not vary across firms or time, so we impose cross-sectional and temporal homogeneity by assuming a single, time-invariant prediction function $f(\cdot)$ that applies to all firms and periods. This restriction allows us to pool the entire panel to learn systematic return predictability and ensures that differences in predicted returns and bound widths reflect cross-sectional variation in characteristics and estimation uncertainty, rather than firm-specific or time-varying model parameters.
In empirical work, one can retrain the model on a rolling basis (e.g., annually) to address temporal nonstationarity while keeping $f(\cdot)$ time-invariant within each estimation window.

Second, we restrict the information set to contemporaneous firm characteristics and macro economic state variables observed at time $t$, excluding lagged returns, realized volatility, and other time-series trading signals. This restriction ensures that the prediction problem targets conditional expected returns rather than return dynamics, and that the bound adjustment captures asset-specific estimation uncertainty rather than volatility timing or momentum effects.

\subsection{Machine Learning Models}

We consider a broad set of ML models that span a wide range of inductive biases, functional flexibility, and estimation stability. This menu closely follows the canonical set analyzed by \cite{GuKellyXiu20}, augmented with modern boosted-tree methods. The goal is not to identify a single ``best'' ML model, but to evaluate whether the proposed uncertainty-adjusted sorting rule delivers robust economic gains across model classes that differ sharply in how they trade off bias, variance, and approximation error.

Greater functional flexibility increases a model's ability to approximate complex conditional mean functions, but it does so at the cost of higher estimation error in finite samples. As a result, while flexible models may improve point predictions on average, they also generate less stable cross-sectional rankings, as small perturbations in the data can lead to large changes in predicted returns. In contrast, more stable and strongly regularized models, such as Elastic Net and PCR, produce concentrated residual distributions and relatively tight uncertainty-adjusted bounds, reflecting more reliable---but less expressive---estimates of expected returns. Highly flexible models, including deep neural networks and boosted tree methods, exhibit more dispersed and heterogeneous residuals, leading to wider bounds and greater cross-sectional uncertainty. This tradeoff highlights the economic role of uncertainty-adjusted sorting: to discipline model flexibility by accounting for estimation risk when translating forecasts into portfolio positions.

Below we briefly introduce the ML models used in this paper and refer to \cite{HastieTibshiraniFriedman09} and \cite{GoodfellowBengioCourville16} for further details and additional models.

\subsubsection{Penalized Linear Model: Elastic Net (ENet).}

Elastic Net is a linear model designed for high-dimensional return-prediction settings, where firm characteristics are numerous, noisy, and strongly correlated \citep{ZouHastie05}.
The model approximates conditional expected returns as a linear function of characteristics, but imposes explicit regulation on coefficient estimates to limit overfitting and improve out-of-sample stability.

This regularization operates through two complementary penalty terms, the $\ell_{1}$ and $\ell_{2}$ penalties. Elastic Net therefore generalizes LASSO \citep{Tibshirani96} and ridge regression \citep{HoerlKennard70}, which use these penalties, respectively.
The $\ell_{1}$ component encourages sparsity by effectively discarding weak or redundant predictors, while the $\ell_{2}$ component shrinks coefficient magnitudes smoothly toward zero, stabilizing estimation in the presence of multicollinearity. Together, these mechanisms prevent extreme coefficient values and reduce sensitivity to sampling variation, a key concern in empirical asset pricing.

Such penalized linear models substantially outperform ordinary least squares when the number of predictors is large relative to the effective time-series sample size. At the same time, the linear structure of Elastic Net limits its ability to capture nonlinearities and higher-order interactions among characteristics. In our setting, Elastic Net therefore serves as a disciplined benchmark with strong estimation stability but limited functional flexibility, allowing us to assess whether uncertainty-adjusted sorting delivers incremental value even when point predictions are already carefully regularized.

\subsubsection{Dimension Reduction Model: Principal Components Regression (PCR).}

Principal Components Regression addresses high dimensionality by summarizing the cross section of firm characteristics into a small number of latent factors \citep{StockWatson02}. Rather than selecting individual predictors, PCR constructs orthogonal linear combinations of characteristics that capture the largest share of cross-sectional variation, and then uses these components to forecast returns.

From an econometric perspective, PCR imposes strong regularization by restricting predictive information to a low-dimensional subspace. This compression substantially reduces estimation noise and multicollinearity, yielding stable coefficient estimates and reliable out-of-sample performance in settings where individual predictors are weak and highly correlated. However, because the components are constructed without reference to future returns, PCR may discard low-variance directions that nonetheless contain economically relevant predictive signals.

In our framework, PCR represents a highly stable but relatively inflexible benchmark. Its strong dimension reduction produces concentrated residual distributions and narrow prediction bounds, making it useful for assessing whether uncertainty-adjusted sorting adds value even when predictions are already smooth and resistant to overfitting.

\subsubsection{Dimension Reduction Model: Partial Least Squares (PLS).}

Partial Least Squares also reduces dimensionality through factor construction, but differs from PCR in that it explicitly incorporates the forecasting objective into the dimension-reduction step. PLS components are chosen to maximize their covariance with future returns, thereby prioritizing predictors that exhibit stronger marginal predictive content \citep{deJong93}.

This design allows PLS to strike a balance between noise reduction and predictive relevance. Relative to PCR, PLS typically retains more economically informative variation, while still imposing substantial regularization through low-dimensional factor structure. As a result, PLS often improves predictive accuracy when signals are weak but systematic, without sacrificing estimation stability.

Within our setting, PLS serves as an intermediate case between highly stable linear benchmarks and more flexible nonlinear models. Its residual distributions are generally more dispersed than those of PCR, reflecting greater responsiveness to predictive signals, yet remain substantially more concentrated than those generated by highly flexible algorithms. This positioning makes PLS informative for evaluating how uncertainty-adjusted sorting performs as estimation uncertainty increases in a controlled manner.

\subsubsection{Tree-Based Ensemble Method: Random Forest (RF).}

Random Forest is a nonlinear ensemble method that captures complex interactions among firm characteristics by averaging a large number of regression trees. Each tree partitions the predictor space into regions defined by recursive binary splits and assigns predictions as local averages within those regions \citep{Breiman01}.

To mitigate the instability of individual trees, Random Forest introduces randomness through bootstrap resampling and random feature selection at each split. These mechanisms reduce variance by decorrelating trees and averaging across many alternative partitions of the data. As shown in \cite{GuKellyXiu20}, this approach allows Random Forests to capture nonlinear relationships while maintaining reasonable out-of-sample robustness.

Despite these stabilizing features, the piecewise-constant nature of tree-based predictions can generate sensitivity near partition boundaries, particularly in sparse regions of the predictor space. In our context, this translates into heterogeneous and asset-specific residual distributions. Random Forest therefore provides a natural setting in which predictive uncertainty varies meaningfully across stocks, making it well suited for evaluating the economic role of uncertainty-adjusted sorting in moderating ranking instability induced by nonlinear interactions.

\subsubsection{Tree-Based Ensemble Method: Gradient Boosting.}

Gradient boosting constructs an additive model of weak learners, typically shallow regression trees, trained sequentially to correct the residuals of earlier learners. Each new tree targets observations the current ensemble predicts poorly, improving fit while controlling complexity through shrinkage and early stopping.

XGBoost (Extreme Gradient Boosting) is a modern, highly optimized implementation of gradient boosting that extends standard boosted regression trees along several dimensions  \citep{ChenGuestrin16}. First, it introduces second-order (Newton-style) optimization of the loss function, improving convergence speed and numerical stability. Second, it incorporates explicit regularization penalties on tree complexity, including penalties on the number of leaves and leaf weights, which directly control model variance. Third, XGBoost supports column subsampling and row subsampling, further enhancing robustness and reducing overfitting.

Relative to traditional gradient boosted regression trees (GBRT) used in \citep{GuKellyXiu20}, XGBoost typically achieves superior performance in noisy prediction environments due to its aggressive regularization and efficient handling of sparse, high-dimensional inputs. These features make XGBoost particularly suitable for asset-pricing data, where signal-to-noise ratios are low and predictor interactions are complex.

From the perspective of our uncertainty-adjusted sorting framework, XGBoost represents a distinctive and informative test case. Its sequential residual-fitting architecture generates rich, asset-specific, and time-varying residual distributions, creating substantial heterogeneity in predictive uncertainty across stocks. At the same time, its regularization mechanisms limit extreme instability, allowing us to assess whether uncertainty-adjusted sorting delivers incremental value even for models that already achieve an effective balance between flexibility and stability. Consistent with this interpretation, uncertainty-adjusted sorting produces smaller---but still statistically significant---performance gains for XGBoost, reinforcing the view that the primary economic role of uncertainty-adjusted sorting is to discipline estimation uncertainty rather than to amplify already strong point predictions.

\subsubsection{Neural Networks (NN1--NN5).}

Neural networks are flexible nonlinear function approximators that model expected returns through layered compositions of affine transformations and nonlinear activation functions. Following \cite{GuKellyXiu20}, we focus on feed-forward multilayer perceptrons \citep{HornikStinchcombeWhite89} with increasing depth and decreasing width.

Shallow networks can be interpreted as nonlinear generalizations of linear factor models, while deeper architectures capture increasingly complex interactions among characteristics. However, the low signal-to-noise ratio and limited effective sample size in asset-pricing applications imply that excessively deep networks tend to overfit. Consequently, predictive performance often peaks at moderate depth.

We therefore consider a fixed set of architectures  with increasing depth, from one to five hidden layers (NN1--NN5).
We train these models by minimizing mean squared prediction error  (MSE) with $\ell_{1}$ regularization and early stopping \citep{GoodfellowBengioCourville16}.
Training uses the Adam optimizer, a stochastic gradient method with adaptive learning rates \citep{KingmaBa15}.
This controlled design allows us to examine how the uncertainty-adjusted sorting rule performs as model flexibility and estimation variance increase.

\subsubsection{Role of Model Diversity in Uncertainty-Adjusted Sorting.}

Across these ML models, inductive biases range from highly stable but inflexible (Elastic Net, PCR) to highly flexible but potentially unstable (XGBoost, deep neural networks). This diversity is intentional. If uncertainty-adjusted sorting improves portfolio performance across such heterogeneous models, the gains cannot be attributed to idiosyncratic features of any single model. Instead, it suggests that incorporating predictive uncertainty into cross-sectional sorting addresses a fundamental economic issue: how estimation risk interacts with portfolio formation, rather than exploiting model-specific artifacts.
Model heterogeneity primarily affects the magnitude and dispersion of estimation uncertainty, while the uncertainty-adjusted sorting rule provides a common, economically interpretable mapping from predictions and uncertainty into portfolio decisions.

\subsubsection{Rolling Estimation and Hyperparameter Selection.}

Each ML model is retrained annually using monthly data.
The rolling scheme is as follows: for the first test year (1992), 1967--1986 is used for training, 1987--1991 for validation, and 1992 is the test year.
Each subsequent year advances a 25-year window: the just-tested year moves into validation, the oldest validation year moves into training, and the oldest training year is dropped.
This rolling design uses only information available prior to each test period, enforces strict out-of-sample evaluation, mitigates  non-stationarity in asset returns, avoids look-ahead bias, and reflects the practical need to update models as new data arrive.

ML models often involve hyperparameters, and their values are critical for performance \citep{FeurerHutter19}. Examples include the regularization weights in Elastic Net and the tree depth and number of trees in random forests. We select hyperparameter values via grid search to minimize validation-set MSE.

\subsection{Uncertainty-Adjusted Bounds}\label{sec:interv-constr}

Given training data of characteristics and realized returns $(x_{i,t}, r_{i,t+1})$ over $t=0,\ldots,T-1$, we train an ML model $\widehat{f}(\cdot)$ to predict the expected return for stock $i$ when a new observation $x_{i,T}$ becomes available.
We seek to quantify uncertainty around the point prediction $\widehat{f}(x_{i,T})$ by constructing an interval $\text{PI}_{i,T}(\alpha)$ that depends on both the training data and the new test point $x_{i,T}$ and contains the realized return $r_{i,T+1}$ with probability at least $\alpha$:
\[\Pr\left(r_{i,T+1} \in \text{PI}_{i,T}(\alpha)\right) \geq \alpha.\]
Conformal prediction \citep{AngelopoulosBates23} provides a principled way to construct such intervals and to quantify predictive uncertainty without parametric assumptions on the data-generating process.

A naive approach is to use in-sample residuals
\[
\widehat{\varepsilon}_{i,t+1} =  r_{i,t+1} - \widehat{f}(x_{i,t}),
\]
to estimate the prediction-error distribution for a new test point. This yields a prediction interval centered at the point forecast $\widehat{f}(x_{i,T+1})$ with half-width equal to the $\alpha$ quantile of the absolute residuals $\{|\widehat{\varepsilon}_{i,1}|,\ldots,|\widehat{\varepsilon}_{i,T}|\}$,
which is the $\lceil \alpha T \rceil$-th smallest value of this set.
However, this approach typically suffers from overfitting: in-sample residuals are smaller than out-of-sample errors, especially for flexible ML models, so the interval is too narrow.

To mitigate overfitting, one can use the split conformal prediction method, which divides the data into two parts. One trains the ML model on one part and compute residuals on the other, so these residuals are out-of-sample errors \citep{LeiGSellRinaldoTibshiraniWasserman18}.
This method trades off model fit against calibration precision.

In empirical asset pricing, the split can leave too little data for training. With monthly data, there are only about 240 observations over two decades per stock, and further extending the window risks nonstationarity. Splitting such data in half yields too few observations to train flexible models, especially deep neural networks. When the training set is small, the fitted model is weak, out-of-sample residuals are large, and the resulting intervals are too wide, causing significant over-coverage.

\subsubsection{Cross-Validated Residuals.}

A common way to address the limits of split conformal prediction is cross-validation \citep{BarberCandesRamdasTibshirani21}. Within each rolling window, we order the data chronologically and partition it into $K$ contiguous, non-overlapping folds, each a consecutive block of time. We do not shuffle observations across folds. Preserving temporal order prevents look-ahead bias and ensures evaluation under realistic information constraints.

The $K$ folds support a time-series cross-validation scheme that separates model training, hyperparameter selection, and out-of-sample residual calculation. For each $k=1,\ldots,K$, we designate the $k$-th fold as the validation set, the $(k+1)$-th (modulo $K$) as the calibration set, and the remaining $K-2$ folds as the training set.
We fit the model on the training folds while selecting hyperparameters using the validation fold, then compute out-of-sample residuals on the calibration fold using the chosen hyperparameters. This yields a sequence of fold-specific models, each estimated strictly prior to the observations on which it is evaluated.

Most importantly for our purposes, this cross-validated rolling procedure produces a disciplined collection of out-of-fold calibration residuals. Because these residuals are generated from predictions that are neither in-sample nor contemporaneously optimized for the calibration observations, they are not mechanically attenuated by overfitting.
As a result, they provide a realistic and economically meaningful characterization of predictive uncertainty, which forms the foundation for the uncertainty-adjusted bound constructed in the subsequent subsection.

\subsubsection{Asset-Level Residual Pooling.}

Unlike standard conformal-prediction settings, where the prediction target is scalar and modeled asset by asset,
here we use a single cross-sectional model $\hat{f}(x)$ to predict returns for many assets simultaneously. As a result, each stock has its own set of out-of-sample residuals obtained via $K$-fold cross-validation. We pool residuals at the asset level, keeping one set per stock rather than pooling across assets.  This pooling strategy lets the prediction-error distribution vary by asset, capturing persistent heterogeneity in estimation uncertainty arising from differences in liquidity, analyst coverage, or fundamental volatility.

Given an residual pool of size $n_i$ for stock $i$, we compute the empirical quantile $\widehat{q}_{i,t}(\alpha)$ as the $\lceil (\alpha n_i) \rceil$-th order statistic.
The uncertainty-adjusted bounds are symmetric around the point prediction $\widehat{f}(x_{i,t})$:
\[
\text{PI}_{i,t}(\alpha) = \big[\, \widehat{f}(x_{i,t}) -  \widehat{q}_{i,t}(\alpha),\; \widehat{f}(x_{i,t}) + \widehat{q}_{i,t}(\alpha) \,\big].
\]
Operationally, these bounds form an empirical quantile band around the ML point prediction. They do not rely on parametric assumptions or asymptotic approximations. Instead, they use the realized dispersion of historical out-of-sample prediction errors to quantify asset-specific predictive uncertainty. These bounds provide the key input to the uncertainty-adjusted sorting rule used in the portfolio construction stage.

\begin{figure}[ht]
\begin{center}
    \caption{Flowchart of Asset-Specific Residual Pooling and Uncertainty-Adjusted Bounds Construction Under a 25-Year Rolling Window} \label{fig:flowchart}
    \includegraphics{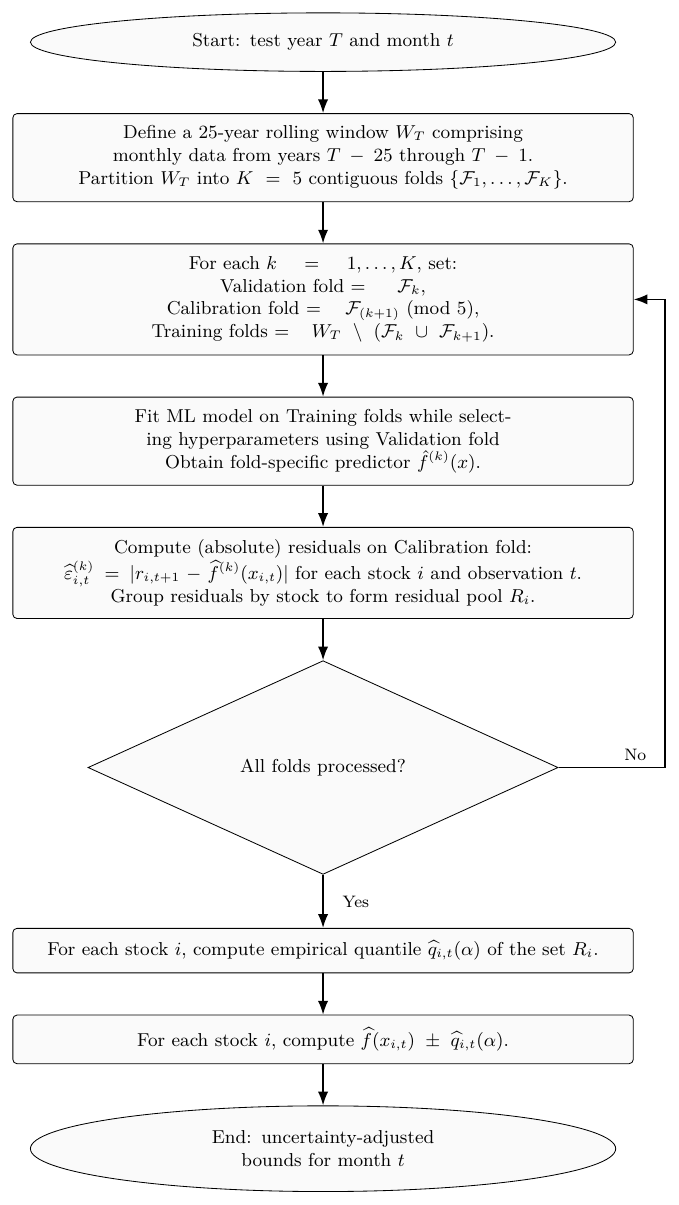}
\end{center}
\end{figure}

The above rolling, time-ordered cross-validation procedure for constructing uncertainty-adjusted bounds is summarized in the flowchart Figure~\ref{fig:flowchart}.

To build intuition for the uncertainty-adjustment framework, Figure~\ref{fig:prediction-illus} provides a stock-level illustration of point predictions and their associated uncertainty-adjusted bounds. The figure plots the predicted next-month excess return for a randomly selected stock, together with the 1\%, 5\% and 10\% quantile-level bounds constructed from historical residual distributions.

Two features are immediately apparent. First, predictive uncertainty is both substantial and time-varying: bound widths expand during periods of heightened volatility and contract during calmer periods. Second, extreme point predictions are often accompanied by wide uncertainty-adjusted bounds, indicating lower reliability precisely when predicted returns appear most attractive or most adverse.

This observation motivates the central idea of the paper. Point predictions alone treat all signals uniformly, regardless of their underlying uncertainty.
Uncertainty-adjusted sorting, by contrast, explicitly accounts for heterogeneity in prediction reliability across assets and over time.
By incorporating this information into portfolio sorting---using upper bounds for long positions and lower bounds for short positions---the uncertainty-adjusted strategy selectively downweights noisy extreme predictions while preserving economically meaningful signals.
The subsequent empirical analysis examines whether this uncertainty-adjusted ranking mechanism improves portfolio performance in the cross section.

\subsection{Portfolio Construction}\label{sec:port-constr}

We now describe how point predictions and uncertainty-adjusted bounds are translated into cross-sectional portfolio sorts.

\subsubsection{Point-Prediction Portfolio Sorting.}

In each month $t$, we generate one-month-ahead return predictions for all stocks in the cross section using the trained ML model. We then sort stocks by predicted returns and assign them to ten deciles. The baseline long–short strategy goes equal weight long in the top decile (decile 10) and equal weight short in the bottom decile (decile 1). This sorting rule follows the empirical asset-pricing literature \citep{GuKellyXiu20} and serves as the benchmark for evaluating uncertainty-adjusted sorting strategies.

\subsubsection{Uncertainty-Adjusted Sorting.}

Let $S_{i,t}^{\text{long}}(\alpha) $ and $S_{i,t}^{\text{short}}(\alpha)$ be the uncertainty-adjusted bounds defined in Section~\ref{sec:interv-constr}:
\[
S_{i,t}^{\text{long}}(\alpha)  = \widehat{f}(x_{i,t}) + \widehat{q}_{i,t}(\alpha)
\quad \mbox{ and } \quad
S_{i,t}^{\text{short}}(\alpha) = \widehat{f}(x_{i,t}) - \widehat{q}_{i,t}(\alpha).
\]

Uncertainty-adjusted portfolio construction modifies the sorting signal by explicitly accounting for predictive uncertainty.
For the long leg, we sort stocks by the upper bound $S_{i,t}^{\text{long}}(\alpha)$; for the short leg, we sort by the lower bound $S_{i,t}^{\text{short}}(\alpha)$.
As in the point-prediction sorting rule, stocks are sorted into deciles.
The strategy goes long the top decile under the upper-bound sorting and short the bottom decile under the lower-bound sorting, yielding a long–short spread.
Equivalently, the spread arises from sorting on two utility-like scores that adjust the point prediction using asset-specific predictive dispersion.

The long score $S_{i,t}^{\text{long}}$ can be interpreted as an optimistic but uncertainty-aware payoff: it rewards a high point prediction and, conditional on a given $\widehat{\mu}_{i,t}$, assigns a higher score to assets whose historical prediction-error distribution allows for larger upside realizations under adverse forecast errors. Conversely, the short score $S_{i,t}^{\text{short}}$ represents a pessimistic but uncertainty-aware payoff: it penalizes assets whose predicted returns remain low even after accounting for favorable prediction-error realizations.

These scores admit a natural economic interpretation. Among stocks with similar point predictions, the long leg favors assets whose realized returns have historically exhibited greater upside resilience under adverse estimation errors, while the short leg targets assets whose low returns persist even when predictions err in a favorable direction.

Importantly, this construction makes clear that the strategy is not a volatility bet. Stocks with high unconditional return volatility are not mechanically favored or penalized. We do not sort on realized return volatility or prediction dispersion per se. Instead, the sorting signal is adjusted using model- and stock-specific predictive dispersion, constructed from out-of-fold calibration residuals produced by the same ML pipeline. The uncertainty-adjusted bound adjustment therefore operates through ranking stability and tail-selection discipline, not by mechanically loading on high-volatility stocks.

A pure volatility-based strategy would tilt toward assets with large unconditional variance regardless of the prediction. In contrast, the adjustment term $\widehat{q}_{i,t}(\alpha)$ penalizes or rewards assets only when the mapping from characteristics to returns is empirically unreliable for a given stock--model pair. Economically, the mechanism operates through an estimation-risk channel: it improves the probability that extreme long and short selections reflect genuine differences in conditional expected returns rather than ranking noise induced by estimation error.

\subsection{Performance Evaluation}

Portfolio performance is evaluated using standard asset-pricing metrics. We report annualized returns, annualized volatility, Sharpe ratios, and Newey--West-adjusted $t$-statistics. For long--short portfolios, we additionally report monthly Fama--French three-factor and five-factor alphas and their corresponding Newey--West $t$-statistics.

For long-only benchmark-relative portfolios, defined as portfolio returns in excess of a market index, factor-adjusted alphas are not computed under the current table design. All performance statistics are computed consistently across point-prediction and uncertainty-adjusted portfolios to isolate the economic effect of uncertainty-adjusted sorting.

\section{An Empirical Study of U.S. Equities}\label{sec:results}

This section presents the empirical results of the uncertainty-adjustment asset pricing framework. We proceed in four steps. First, we document baseline predictive performance using point predictions in both long--short and benchmark-relative long-only portfolios. Second, we evaluate the economic viability of these strategies under realistic transaction costs. Third, we compare point-prediction--based portfolio construction with uncertainty-adjusted portfolio construction across models. Finally, we investigate the firm-level and macroeconomic drivers of performance improvements induced by uncertainty-adjusted sorting.

\subsection{Data}

\begin{table}[ht]
\TABLE{Data Blocks and Key Processing Steps \label{tab:data_blocks}}
{
\begin{tabular}
{@{}lllll@{}}
\toprule
\textbf{Block} & \textbf{Source} & \textbf{Frequency} & \textbf{Span} &  \textbf{Key Processing Steps} \\
\midrule
Stock returns & CRSP & Monthly & 1967--2016 &  \makecell[cl]{Compute next-month excess \\ returns using the 1M T-bill rate.} \\[2ex]
\makecell[cl]{46 Firm \\ characteristics} & \makecell[cl]{CRSP/Compustat; \\ CRSP} & \makecell[cl]{Monthly / \\ Annually}  & 1967--2016 & \makecell[cl]{Annual variables updated at end of \\ June;  monthly variables updated \\ at month-end; cross-sectional \\  quantile transform each month.} \\[5ex]
\makecell[cl]{178 Macroeconomic \\ variables} & \makecell[cl]{FRED-MD; medians  \\ of  characteristics; \\ \cite{WelchGoyal08}} & Monthly & 1967--2016 &  \makecell[cl]{Apply standard transformations; \\ expand to panel by assigning the \\ same macro vector to all stocks \\  each month.} \\
\bottomrule
\end{tabular}}
{\emph{Note.} The monthly sample covers January 1967 to December 2016 and includes the full CRSP common-stock universe. The target variable is the one-month-ahead stock excess return, defined as the CRSP return minus the one-month Treasury bill rate. The predictor set consists of 46 lagged firm characteristics and 178 macroeconomic variables. Firm characteristics include annual accounting variables updated at the end of June and monthly variables measured at month-end. To ensure comparability across firms, firm characteristics are ranked cross-sectionally each month and transformed via a quantile mapping. Macroeconomic predictors are standardized and merged to the stock-level panel by assigning the same macroeconomic vector to all stocks within a given calendar month.}

\end{table}

Table~\ref{tab:data_blocks} summarizes the data blocks used in the empirical analysis and the key preprocessing steps applied prior to model estimation. The sample spans nearly five decades, from January 1967 through December 2016, and covers the full CRSP universe of U.S. common stocks. The forecasting target is the next-month stock excess return, measured relative to the one-month Treasury bill rate, which is standard in the return-predictability and asset-pricing literature.

The predictor set combines rich cross-sectional and time-series information. Firm-level predictors consist of 46 lagged characteristics commonly used in empirical asset pricing, including both annual accounting variables---updated following standard June timing conventions---and monthly characteristics observed at the end of each month. In addition, we include a broad set of 178 macroeconomic predictors that capture aggregate economic and financial conditions.

To mitigate scale differences and ensure stability in cross-sectional comparisons, firm characteristics are transformed using a rank-based procedure: each characteristic is ranked cross-sectionally and mapped to its empirical quantile distribution on a monthly basis. This transformation reduces the influence of outliers and ensures that the predictive signals are comparable across firms and over time. Macroeconomic variables are standardized and merged into the stock-level panel by assigning the same macroeconomic vector to all firms within a given month, reflecting the common information set available to investors at that point in time.

Together, these data blocks provide a high-dimensional forecasting environment that closely follows the modern ML asset-pricing literature, while the preprocessing steps ensure that both firm-level and macroeconomic predictors are used in a disciplined and economically interpretable manner.

\subsection{Baseline Point-Prediction Portfolio Performance}

Table~\ref{tab:baseline-performance} reports the baseline performance of portfolios constructed using ML point predictions of one-month-ahead stock excess returns. These results establish the benchmark level of economic predictability achieved by modern ML methods before incorporating predictive uncertainty.

\begin{table}[ht]
\TABLE{Baseline Point-Prediction Portfolio Performance \label{tab:baseline-performance}}
{\includegraphics[height=0.85\textheight]{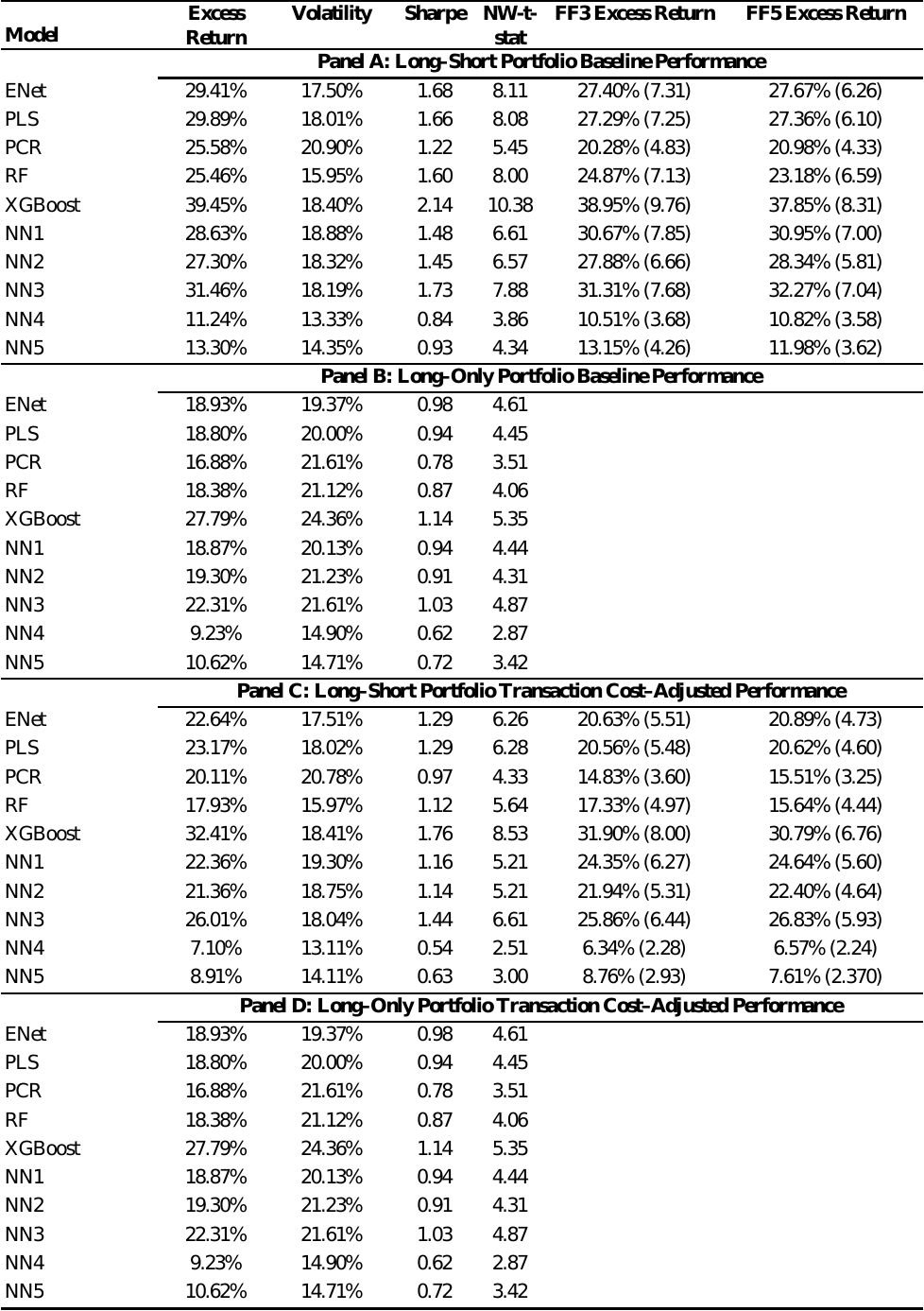}}
{\emph{Note.} This table reports out-of-sample performance (1992–2016) of monthly rebalanced portfolios formed from machine-learning point predictions of next-month stock excess returns. At the end of month $t$, stocks are ranked by predicted returns for month $t+1$. Panel A (Long–Short, gross) forms an equal-weighted long–short decile portfolio: long the top decile and short the bottom decile. Panel B (Long-Only, gross) holds the equal-weighted top-decile portfolio; excess return is measured relative to the S\&P 500 index. Returns and volatility are annualized from monthly data, and Sharpe ratios are annualized accordingly. Panel C (Long–Short, net) reports transaction cost–adjusted performance for the strategies in Panel A. Panel D (Long-Only, net) reports transaction cost–adjusted performance for the strategies in Panel B. Net returns subtract proportional transaction costs of 20 basis points per trade. For long–short portfolios, FF3 and FF5 alphas (monthly) are intercepts from time-series regressions on the Fama–French three- and five-factor models; Newey–West $t$-statistics are reported in parentheses. FF3/FF5 alphas are not reported for long-only portfolios because the Fama–French factors are constructed as long–short portfolios and do not have a long-only analogue.}
\end{table}

\subsubsection{Baseline Long--Short Portfolios.}

Panel A of Table~\ref{tab:baseline-performance} reports the baseline performance of long--short portfolios constructed by ranking stocks on ML point predictions of one-month-ahead returns. Each month, stocks are sorted into deciles based on predicted returns; the portfolio goes long the top decile and short the bottom decile, with equal weighting within each leg. Across most model classes, point-prediction strategies generate economically large and statistically significant return spreads. Annualized long--short returns range from approximately 25\% to nearly 40\%, with Sharpe ratios exceeding 1.2 for the better-performing models. Newey--West adjusted $t$-statistics for excess returns are uniformly large---often above 7---indicating strong statistical significance.

A key finding is the dominant performance of XGBoost. The XGBoost long--short portfolio delivers an annualized return of 39.45\% with annualized volatility of 18.40\%, corresponding to a Sharpe ratio of 2.14. The Newey--West $t$-statistic for the mean excess return is 10.38. Importantly, this performance is not explained by standard risk factors: both FF3 and FF5 alphas are 0.032 per month, with $t$-statistics of 9.76 and 8.31, respectively. These results indicate that XGBoost captures cross-sectional return variation well beyond linear factor structures.

Regularized linear and dimension-reduction models also perform strongly but are clearly dominated by boosted trees. Elastic Net and PLS produce annualized returns close to 30\%, with Sharpe ratios around 1.66--1.68 and FF5 alpha $t$-statistics exceeding 6. PCR, while still profitable, performs noticeably worse, with a Sharpe ratio of 1.22, highlighting the limitations of purely linear dimension-reduction approaches in this setting.

Neural networks exhibit pronounced heterogeneity. Shallow and moderately deep architectures (NN1--NN3) perform well, with Sharpe ratios between 1.45 and 1.73 and economically meaningful factor-adjusted alphas. In contrast, deeper architectures (NN4 and NN5) perform poorly, with Sharpe ratios below 1 and substantially weaker statistical significance. This dispersion foreshadows a central theme of the paper: greater model flexibility can enhance predictive power, but it also increases vulnerability to estimation error.

Overall, Panel A of Table~\ref{tab:baseline-performance}  establishes that ML point predictions contain substantial cross-sectional information and provide a demanding benchmark for any alternative portfolio construction method.

\subsubsection{Baseline Long-Only Portfolios.}

Panel B of Table~\ref{tab:baseline-performance}  reports benchmark-relative long-only performance, where excess returns are measured relative to the S\&P~500 index. This design evaluates whether predictive signals remain economically meaningful in the absence of short-selling or leverage.

The results broadly mirror those in the long--short setting but with attenuated magnitudes, as expected. XGBoost again delivers the strongest performance, generating an annual excess return of 27.79\% (Newey--West $t$-statistic of 5.35) with annualized volatility of 24.36\% and a Sharpe ratio of 1.14. Among neural networks, NN3 performs best, producing an annual excess return of 22.31\% and a Sharpe ratio of 1.03.

Elastic Net, PLS, and Random Forest also deliver economically meaningful benchmark-relative excess returns, with Sharpe ratios close to 1. In contrast, NN4 and NN5 again underperform, with Sharpe ratios below 0.75, indicating that deeper neural-network architectures struggle even in a long-only setting.

Factor-adjusted alphas based on the Fama--French three- and five-factor models are not reported for the long-only portfolios. This is because the Fama--French factors themselves are constructed as long--short portfolios and therefore do not admit a natural long-only analogue. As a result, FF3 and FF5 regressions are only meaningful for dollar-neutral strategies and are appropriately reported only for the long--short portfolios in Panel A of Table~\ref{tab:baseline-performance} .

Taken together, Panels A and B of Table~\ref{tab:baseline-performance}  show that ML point predictions generate economically significant predictability in both dollar-neutral and benchmark-relative portfolios, but that model choice plays a critical role. These baseline results motivate the subsequent analysis, which asks whether incorporating predictive uncertainty can further improve portfolio performance, particularly for models where estimation risk is most pronounced.

\subsubsection{Transaction Cost--Adjusted Long--Short Portfolios}

Panel C of Table~\ref{tab:baseline-performance}  evaluates the robustness of the baseline long--short strategies after incorporating transaction costs. Following Avramov et al.\ (2023), we impose a proportional trading cost of 20 basis points per trade. This assumption is intentionally conservative and is meant to serve as a stress test representative of active equity trading environments.

As expected, transaction costs mechanically reduce returns across all models. However, several important patterns remain. First, XGBoost continues to dominate economically and statistically after costs. The transaction cost--adjusted annualized return remains high at 32.41\%, with a Sharpe ratio of 1.76 and a Newey--West $t$-statistic of 8.53. Factor-adjusted performance remains strong: both FF3 and FF5 alphas are large and highly significant. These results indicate that the profitability of the XGBoost strategy is not driven by ignoring trading frictions.

Second, Elastic Net and PLS remain robust to transaction costs. Both models continue to deliver economically meaningful returns, with Sharpe ratios around 1.29 and Newey--West $t$-statistic above 6. This suggests that their predictive signals are sufficiently stable and persistent to survive aggressive cost assumptions.

Third, neural networks again exhibit pronounced heterogeneity. NN3 remains economically viable after costs, with a Sharpe ratio of 1.44, whereas NN4 and NN5 deteriorate sharply and become economically marginal. This pattern reinforces the interpretation that greater model flexibility can amplify estimation instability and turnover, making performance more vulnerable to trading frictions.

Overall, Panel C of Table~\ref{tab:baseline-performance}  shows that the strongest ML signals remain economically viable even after accounting for substantial transaction costs, while weaker or less stable models are appropriately disciplined once trading frictions are introduced.

\subsubsection{Transaction Cost--Adjusted Long-Only Portfolios}

Panel D of Table~\ref{tab:baseline-performance}  reports transaction cost--adjusted performance for benchmark-relative long-only portfolios. This analysis evaluates whether predictive signals remain economically meaningful in the absence of short-selling once trading frictions are taken into account.

The ranking of models remains largely consistent with the baseline results. XGBoost again delivers the strongest performance, with a post-cost Sharpe ratio of 0.90. Elastic Net and PLS remain positive, though with more modest performance, reflecting the combined effects of transaction costs and the absence of leverage. Random Forest performs comparably but does not dominate.

In contrast, NN4 and NN5 perform the worst, with low Sharpe ratios and weak economic significance, consistent with earlier findings. This confirms that deeper neural-network architectures struggle to generate robust benchmark-relative performance once trading frictions are introduced.

Taken together, Panels C and D of Table~\ref{tab:baseline-performance}  confirm that predictive performance from ML models persists under conservative transaction cost assumptions, but that economic significance is concentrated among a subset of models with relatively stable predictive signals.

\subsection{Uncertainty-Adjusted Sorting Versus Point Prediction}

\afterpage{
\clearpage
\begin{landscape}
\begin{table}[ht]
    \centering
    \caption{Uncertainty-Adjustment-Sorted Portfolio Performance}
    \label{tab:ci_results}
    \begin{threeparttable}
    \includegraphics[width=1.35\textheight]{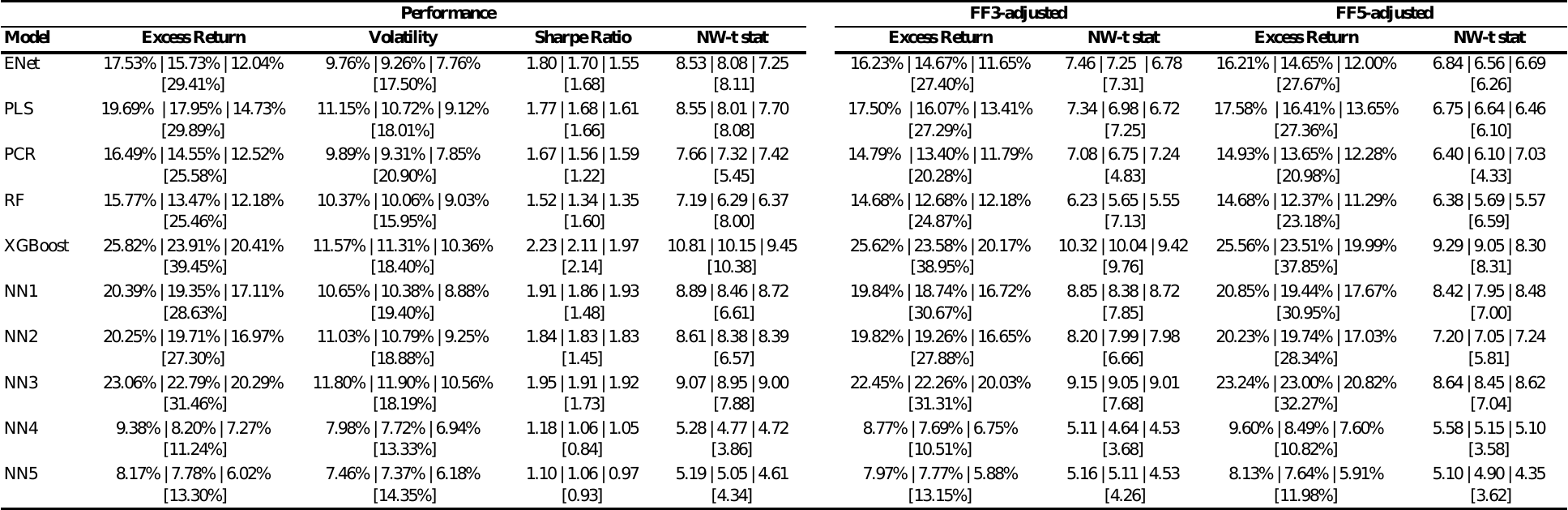}
        \begin{tablenotes}
            \item \scriptsize{\emph{Note.} This table compares uncertainty-adjusted portfolio construction with point-prediction benchmarks. Uncertainty-adjusted-sorting results are reported outside square brackets, while point-prediction results are reported in square brackets. The left (middle, right) bound estimate corresponds to the 1\% (5\%, 10\%) residual-quantile. All results are computed over the out-of-sample period 1992–2016. Uncertainty-adjusted bounds are constructed by combining each model's point prediction with asset-specific residual-quantile half-widths estimated from a rolling window of past out-of-fold prediction errors. Portfolios are formed monthly by ranking stocks on upper bounds for the long leg and lower bounds for the short leg. Returns and volatilities are annualized; Sharpe ratios are reported in annualized terms.}
        \end{tablenotes}
    \end{threeparttable}

\end{table}
\end{landscape}
}

Table~\ref{tab:ci_results} is central to this paper. It directly compares uncertainty-adjusted portfolio construction with conventional point-prediction-based sorting, holding all other aspects of the strategy fixed. Results based on uncertainty-adjusted bounds are reported outside brackets, while point-prediction benchmarks are shown inside brackets. We report results for three quantile levels---1\%, 5\% and 10\% from the left to the middle to the right---which index increasingly aggressive uncertainty adjustments and allow us to study how increasing robustness affects portfolio performance. Three robust findings emerge.

\subsubsection{Sharpe Ratio Improvements Driven by Volatility Reduction.}

A robust pattern across models is that uncertainty-adjusted sorting often reduces average returns relative to point-prediction sorting, but reduces return volatility by an even larger margin. As a result, uncertainty-adjusted performance improves for most models, as reflected in higher Sharpe ratios and larger Newey--West $t$-statistics.

For example, NN1's annualized return declines from 28.63\% under point-prediction sorting to 19.35\% under 5\% uncertainty-adjusted sorting, yet its Sharpe ratio increases from 1.48 to 1.86, and the Newey--West $t$-statistic rises from 6.61 to 8.46. Similar patterns hold for eight out of ten models including other neural-network models, ENet, PLS and PCR. These results indicate that uncertainty-adjusted sorting improves signal-to-noise efficiency not by inflating expected returns, but by reducing volatility, stabilizing portfolio payoffs and mitigating noise-driven extreme-decile selection.

As the adjustment becomes more aggressive to 10\% quantile level, volatility continues to decline, but return attenuation becomes more pronounced. This trade-off highlights the economic role of uncertainty adjustment: stronger penalization increases robustness but can become overly conservative when the underlying signal is already reliable.

\subsubsection{Large Gains for Flexible Models.}

The largest gains from uncertainty-adjusted sorting arise for flexible models, particularly neural networks, which are most exposed to estimation uncertainty. For several neural architectures, Sharpe ratios increase substantially when uncertainty-adjusted bounds are used for ranking, accompanied by marked reductions in volatility and improvements in statistical significance.

Notably, even neural network models that perform poorly under point-prediction ranking exhibit sizable relative improvements under uncertainty-adjusted sorting. This pattern strongly supports the interpretation that uncertainty adjustment disciplines flexible learners by penalizing unreliable extreme predictions, rather than suppressing predictive signal uniformly.

\subsubsection{Limited or Diminishing Gains for Strong Models.}

For models that already perform exceptionally well under point-prediction ranking—most notably XGBoost—the incremental benefits of uncertainty-adjusted sorting are more limited. Moderate uncertainty adjustment yields small improvements in Sharpe ratios, primarily through volatility reduction. However, under more aggressive adjustment, these gains diminish and may even reverse.

This pattern is consistent with diminishing marginal returns to robustness when baseline predictions are already stable. When point predictions are accurate and well-ordered, aggressive uncertainty penalization can unnecessarily compress return spreads and attenuate genuine signal.

\subsubsection{Scaled Net-Value Paths.}

\begin{figure}[ht]
    \begin{center}
        \caption{Scaled Net-Value Paths: Uncertainty-Adjusted Sorting versus Point-Prediction Sorting \label{fig:scaled-net-value}}
        \includegraphics[width=\textwidth]{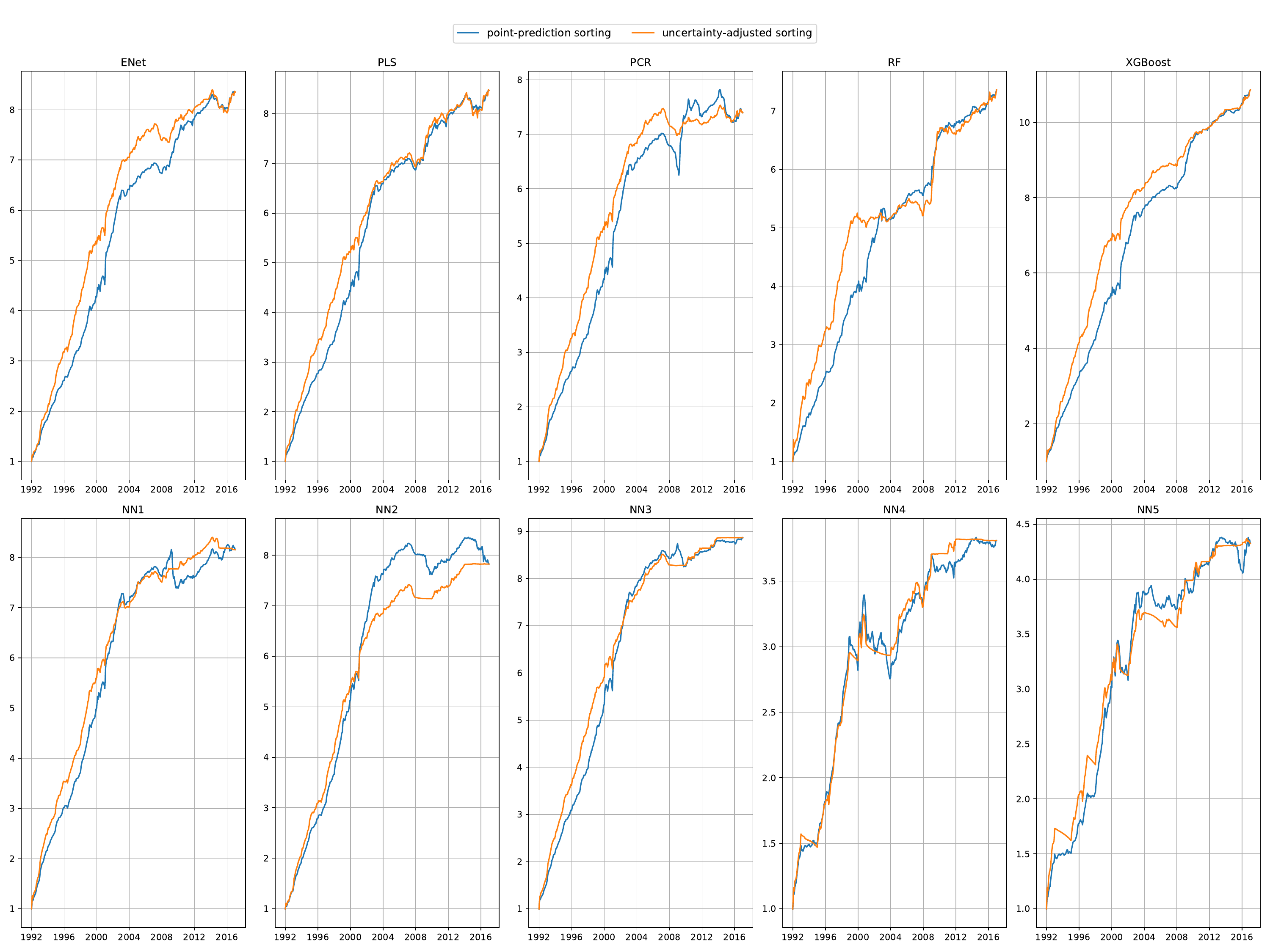}
    \end{center}
    \begin{tablenotes}
        \item \scriptsize{\emph{Note.} {This figure compares the uncertainty-adjusted cumulative net value curves of long–short portfolios constructed from (i) point predictions from ten models and (ii) the 5\% quantile uncertainty-adjusted bounds reported in Table~\ref{tab:ci_results}. Because our primary evaluation metric is the Sharpe ratio, in each panel the portfolio formed using point predictions is treated as the benchmark, and the portfolio formed using uncertainty-adjusted bounds is levered so that the two portfolios have the same annualized return. This normalization facilitates a more direct visual comparison of their performance trajectories.}}
    \end{tablenotes}
\end{figure}

Figure~\ref{fig:scaled-net-value} provides a visual comparison between uncertainty-adjusted portfolios and point-prediction-based portfolios, focusing explicitly on return volatility and path smoothness, rather than differences in terminal performance.

For each ML model, we compare two monthly rebalanced long-short portfolios over the period 1992--2016: (i) a baseline portfolio constructed using point-prediction sorting, and (ii) an uncertainty-adjusted portfolio constructed using bounds corresponding to a moderate 5\% quantile level.

To facilitate a clean comparison of volatility dynamics, both portfolios are initialized with a net asset value of one in January 1992. Because uncertainty-adjusted portfolios typically generate lower cumulative returns than their point-prediction counterparts (as documented in Table~\ref{tab:ci_results}), we apply a simple scaling transformation to the uncertainty-adjusted portfolio's net-value path. Specifically, the uncertainty-adjusted portfolio's monthly net values are multiplied by a constant factor so that its terminal net asset value in December 2016 matches that of the corresponding point-prediction portfolio. In reality, this scaling is equivalent to applying financial leverage in trading. This scaling preserves the entire time-series shape and relative month-to-month fluctuations of the uncertainty-adjusted portfolio, while eliminating differences in final wealth levels. Importantly, this normalization does not affect volatility, Sharpe ratios, drawdowns, any statistical inference, or any portfolio performance metrics reported elsewhere in the paper; it merely aligns terminal values to enable a direct visual comparison of path smoothness.

Across most of the ten models, the uncertainty-adjusted portfolios (orange lines) exhibit noticeably smoother net-value paths than the point-prediction portfolios (blue lines). The reduction in high-frequency fluctuations and drawdown severity is particularly pronounced for flexible models such as neural networks, where estimation uncertainty is largest. Figure~\ref{fig:scaled-net-value} thus provides a graphical counterpart to Table~\ref{tab:ci_results}, illustrating that improvements in risk-adjusted performance arise primarily through volatility stabilization rather than higher mean returns.

\subsubsection{Economic Interpretation: Robust Sorting Rather Than Return Inflation.}

Taken together, Table~\ref{tab:ci_results} shows that uncertainty-adjusted sorting acts as a robustness-enhancing transformation of the sorting rule, rather than a mechanism for inflating mean returns. Improvements in uncertainty-adjusted performance arise primarily through variance reduction, ranking stability, and disciplined tail selection.

The comparison across 1\%, 5\% and 10\% quantile levels further clarifies an important economic trade-off underlying uncertainty-adjusted sorting. Incorporating uncertainty-adjusted bounds effectively trims extreme predictions, reducing both expected returns and return volatility by discarding unstable extreme-decile selections. However, the width of the bound is crucial. When the bound is moderately narrow—corresponding to lower quantile levels that remain close to the point prediction—it filters out the most extreme and noisy forecasts while largely preserving informative signal. In this case, volatility declines more than mean returns, leading to higher Sharpe ratios and visibly smoother return paths, as illustrated in Figure~\ref{fig:scaled-net-value}. When the bound becomes wider, the adjustment removes a broader set of return realizations, including economically meaningful signal. Although volatility continues to decline, the associated reduction in expected returns dominates, resulting in lower risk-adjusted performance.

\subsubsection{Choice of Bound Width.}

The effectiveness of uncertainty-adjusted sorting depends critically on the width of the uncertainty-adjusted bounds. Bound width determines the extent to which extreme predictions are filtered out of the cross-sectional ranking and therefore governs the trade-off between noise reduction and signal preservation.

Narrow bounds introduce only limited adjustment relative to point-prediction ranking and primarily remove the most unstable extreme forecasts. This selective trimming stabilizes portfolio construction by mitigating noise-driven rank reversals while retaining most of the predictive signal. As a result, volatility declines disproportionately relative to expected returns, leading to improvements in risk-adjusted performance.

In contrast, excessively wide bounds impose a much stronger adjustment that compresses the cross-sectional dispersion of predicted returns. While this further reduces volatility, it also eliminates economically meaningful variation in expected returns. Once the bound becomes sufficiently wide, the marginal reduction in volatility is outweighed by the loss of return dispersion, causing Sharpe ratios to deteriorate.

These opposing forces generate an interior optimum in bound width rather than a monotonic relationship between adjustment strength and performance. The location of this optimum varies systematically with model flexibility: models with greater estimation uncertainty benefit more from moderate adjustment, whereas already stable models exhibit smaller gains and are more sensitive to over-adjustment. Overall, effective uncertainty-adjusted sorting requires bounds that are tight enough to discipline unreliable extremes, yet not so wide that the ranking signal itself is diluted.

\subsection{Drivers of Uncertainty-Adjusted Sorting Performance} \label{sec:rank-improvement}

\begin{table}[ht]
    \TABLE{Variable Definitions \label{tab:variables}}
    {\begin{tabular}{@{}lll@{}}
\toprule
\textbf{Variable} & \textbf{Description} & \textbf{Notes} \\
\midrule
SIZE & Firm size & Cross-sectional z-score \\
BEME & Book-to-market ratio & Cross-sectional z-score \\
OP & Operating profitability & Cross-sectional z-score \\
INV & Capital investment & Cross-sectional z-score \\
MKT\_RF & Market return minus risk-free rate & Cross-sectional z-score \\
CUR & Squared change in the civilian unemployment rate & Time-series z-score \\
M1 & Squared change in the natural logarithm of the M1 monetary stock & Time-series z-score \\
M2 & Squared change in the natural logarithm of the M2 monetary stock & Time-series z-score \\
SP500 & Change in the natural logarithm of the S\&P 500 Index & Time-series z-score \\
FEDF & Change in the federal funds rate & Time-series z-score \\
CPI & Squared change in the natural logarithm of the consumer price index & Time-series z-score \\
VXO & Volatility index & Time-series z-score \\
\bottomrule
\end{tabular}
}
{\emph{Note.} This table defines the firm-level and macroeconomic variables used in the driver regressions reported in Table 8. Firm characteristics include size (SIZE), book-to-market ratio (BEME), operating profitability (OP), and investment (INV), constructed following standard Fama--French definitions. All firm characteristics are standardized cross-sectionally within each month. Macroeconomic variables---including market excess return (SP500), federal funds rate (FEDF), inflation (CPI), and market volatility (VXO)---are standardized as time-series z-scores and merged to the stock panel by assigning the same macro realization to all firms within a given month. Interaction terms between firm characteristics and macroeconomic variables are constructed to capture state-dependent heterogeneity in ranking improvements.}

\end{table}

Tables~\ref{tab:variables} and~\ref{tab:regression_results} investigate \emph{why} uncertainty-adjusted sorting improves portfolio performance relative to point-prediction sorting, as documented in Table~\ref{tab:ci_results} and Figure~\ref{fig:scaled-net-value}. Rather than treating uncertainty-adjusted sorting as a black-box adjustment, we examine whether the induced ranking changes are systematically related to firm fundamentals and macroeconomic states---consistent with an estimation-uncertainty channel.

Table~\ref{tab:variables} defines the firm-level characteristics and macroeconomic variables used in the driver regressions. Firm characteristics capture cross-sectional heterogeneity in size, valuation, profitability, and investment, while macro variables proxy for aggregate market conditions, monetary policy, inflation, and uncertainty. Interaction terms allow the effect of uncertainty-adjusted sorting to vary across macroeconomic regimes.

Table~\ref{tab:regression_results} reports panel fixed-effects regressions that relate \emph{ranking improvements} of the ENet model\footnote{Although we only present the results for the ENet model, the significance and the sign direction of many coefficients' estimates are the same for many other ML models. The other models' results are presented in the e-companion.}---defined as the change in cross-sectional rank when moving from point-prediction sorting to uncertainty-adjusted sorting---to these firm-level and macro variables. Specifically, to quantify how firm characteristics and macroeconomic factors drive the improvement of uncertainty-adjusted sorting, we estimate the following fixed-effects regressions separately for the long-leg and short-leg ranking improvements. By construction, a higher value of $\mathrm{rank}$ indicates a more favorable position in the cross-sectional ordering, implying a higher likelihood of being selected into the portfolio’s long leg.

Long-leg ranking improvement:
\[
\Delta \mathrm{rank}_{i,t}^{\mathrm{upper}} = \mathrm{rank}_{i,t}^{\mathrm{upper}} - \mathrm{rank}_{i,t}^{\mathrm{point}}
\]

Short-leg ranking improvement:
\[
\Delta \mathrm{rank}_{i,t}^{\mathrm{lower}} = \mathrm{rank}_{i,t}^{\mathrm{point}} - \mathrm{rank}_{i,t}^{\mathrm{lower}}
\]

The regression models are:
\begin{gather*}
\Delta \mathrm{rank}_{i,t}^{\mathrm{upper}} = \BFbeta_{1}^{\mathrm{upper}} \BFX_{i,t} + \BFbeta_{2}^{\mathrm{upper}} \BFZ_t +  \BFbeta_{3}^{\mathrm{upper}} (\BFX_{i,t} \circ \BFZ_t) + \gamma_i + \varepsilon_{i,t}^{\mathrm{upper}},   \\
\Delta \mathrm{rank}_{i,t}^{\mathrm{lower}} = \BFbeta_{1}^{\mathrm{lower}} \BFX_{i,t} + \BFbeta_{2}^{\mathrm{lower}} \BFZ_t  +    \BFbeta_{3}^{\mathrm{lower}} (\BFX_{i,t} \circ \BFZ_t) + \gamma_i + \varepsilon_{i,t}^{\mathrm{lower}}.
\end{gather*}
Here $\gamma_i$ denotes firm fixed effects that absorb time-invariant firm heterogeneity; $\BFX_{i,t}$ is the vector of standardized firm characteristics, $\BFZ_t$ is the vector of standardized macroeconomic state variables (shared by all firms in month $t$), and $\BFX_{i,t} \circ \BFZ_t$ denotes element-wise interactions that allow the effect of firm characteristics on ranking improvements to vary with macro states; $\varepsilon_{i,t}^{\mathrm{upper}}$ and $\varepsilon_{i,t}^{\mathrm{lower}}$ are idiosyncratic errors.

\afterpage{
\clearpage
\begin{landscape}
\begin{table}[ht]
    \centering
    \caption{Regression Evidence on Ranking Improvements from Uncertainty-Adjusted Sorting: ENet}
    \label{tab:regression_results}
    \begin{threeparttable}
    \includegraphics[width=1.35\textheight]{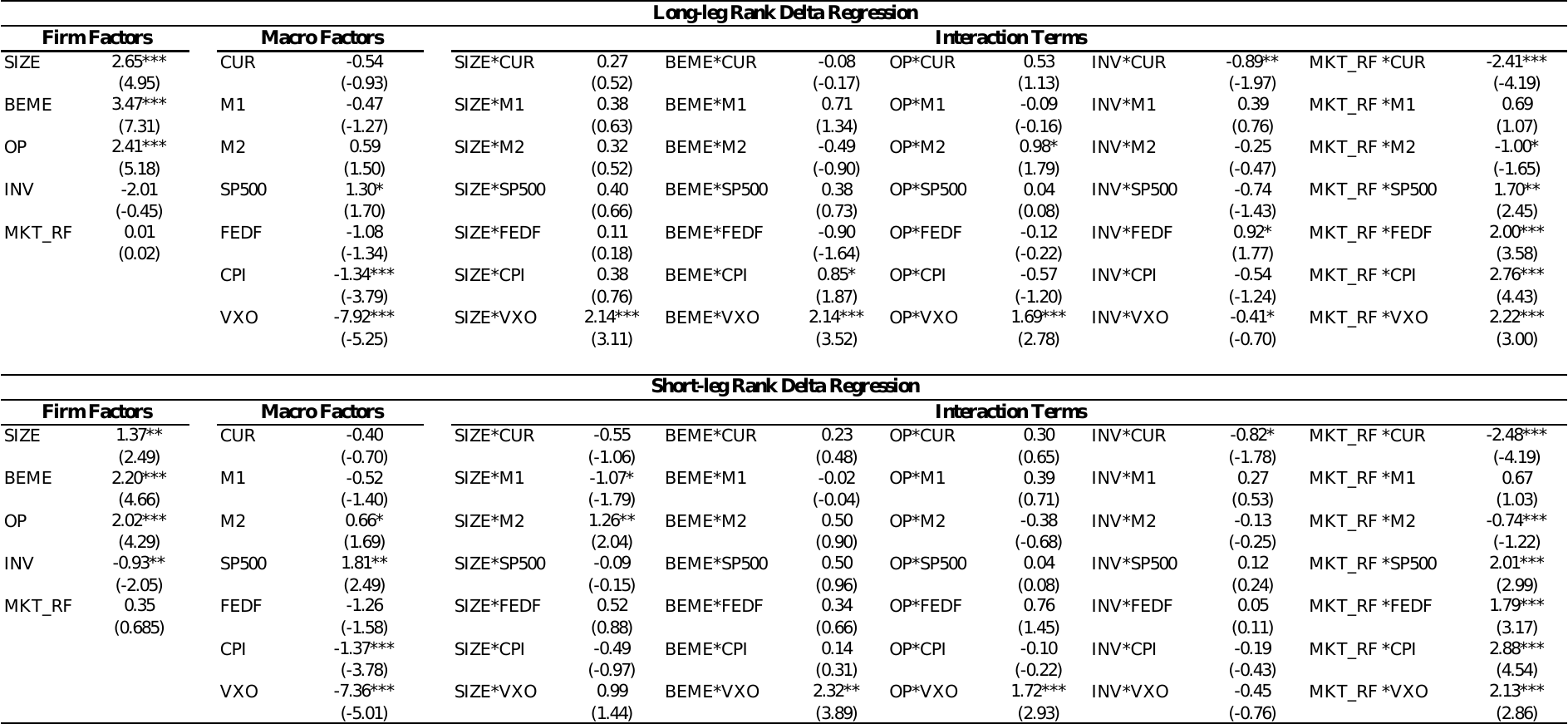}
        \begin{tablenotes}
            \item \scriptsize{\emph{Note.} This table reports panel fixed-effects regressions examining when uncertainty-adjusted sorting improves stock selection relative to point-prediction ranking, using the Elastic Net model with 5\% uncertainty-adjusted bounds. The sample covers the 1992–2016 out-of-sample test period. The dependent variable is the long-leg's and short-leg's ranking improvements defined as follows: $\Delta \mathrm{rank}_{i,t}^{\mathrm{upper}} = \mathrm{rank}_{i,t}^{\mathrm{upper}} - \mathrm{rank}_{i,t}^{\mathrm{point}}$ and $\Delta \mathrm{rank}_{i,t}^{\mathrm{lower}} = \mathrm{rank}_{i,t}^{\mathrm{point}} - \mathrm{rank}_{i,t}^{\mathrm{lower}}$. Explanatory variables include five firm characteristics, seven macroeconomic variables, and their thirty-five interactions. Firm fixed effects are included. Reported coefficients are accompanied by z-statistics in parentheses, based on heteroskedasticity-robust standard errors.}
        \end{tablenotes}
    \end{threeparttable}

\end{table}
\end{landscape}
}

\subsubsection{Firm-Level Heterogeneity in Ranking Improvements.}

As shown in Table~\ref{tab:regression_results}, firm characteristics matter. Ranking improvements are positively associated with firm size (SIZE) and book-to-market (BEME) on both the long and short legs. Economically, this indicates that uncertainty-adjusted sorting reshapes rankings more strongly among larger and value-oriented firms. These firms tend to have richer information environments and more stable fundamentals, making estimation uncertainty more structured rather than purely idiosyncratic.

Also, operating profitability (OP) loads positively and significantly on long-leg rank changes, indicating that uncertainty adjustment affects the long-side ordering more strongly among more profitable firms (given our rank convention, a positive coefficient corresponds to upward movement in the long-side rank).

\subsubsection{State Dependence and the Role of Aggregate Market Uncertainty.}

A potential concern is how macroeconomic factors, such as aggregate uncertainty, proxied by the volatility index (VXO), can affect cross-sectional ranking changes when they take the same value for all stocks at a given point in time. This concern is resolved by recognizing that stock-specific estimation uncertainty does not enter the regressions explicitly on the right-hand side, but is instead embedded in the construction of the dependent variable, $\Delta \mathrm{rank}_{i,t}$.

By definition, $\Delta \mathrm{rank}_{i,t}$ measures the difference between uncertainty-adjusted ranking and point-prediction ranking for a given stock. The uncertainty-adjusted upper and lower bounds are constructed using stock-specific prediction residuals, and therefore reflect heterogeneity in estimation uncertainty across assets. As a result, even though VXO is common across stocks within a given month, the magnitude and direction of ranking adjustments differ across firms because they are mediated through heterogeneous, asset-level uncertainty embedded in the ranking measure itself.

The significantly negative coefficient on VXO thus captures how aggregate market uncertainty scales the overall aggressiveness of uncertainty adjustment, rather than inducing a uniform shift in rankings. When market-wide uncertainty is high, uncertainty-adjusted bounds widen mechanically for all stocks. This induces a more conservative sorting rule, compressing extreme rankings and reducing the likelihood that any stock—long or short—is placed in the most extreme deciles unless its signal is sufficiently robust. Consequently, the difference between uncertainty-adjusted ranks and point-prediction ranks tends to shrink, leading to smaller $\Delta \mathrm{rank}_{i,t}$ on average.

Economically, this result indicates that uncertainty-adjusted sorting becomes more cautious during periods of elevated aggregate volatility. High VXO environments amplify estimation risk across the cross-section, increasing the cost of ranking errors driven by noisy predictions. The uncertainty-adjusted procedure responds endogenously by attenuating extreme rank changes, thereby reducing exposure to unstable signals. This mechanism explains why the VXO coefficient is negative: higher market uncertainty dampens the extent to which uncertainty-adjusted bounds reshape rankings relative to point predictions, even though the underlying stock-level uncertainty remains heterogeneous.

Importantly, this effect does not imply that all stocks’ ranks move synchronously. Rather, VXO governs the intensity of the uncertainty-adjustment mechanism, while cross-sectional heterogeneity in prediction reliability determines which stocks experience larger or smaller ranking revisions. Aggregate uncertainty therefore operates as a state variable that modulates the strength of uncertainty-adjusted sorting, not as a determinant of relative ranking by itself.

\subsubsection{Interaction Effects and Economic Interpretation.}

The interaction terms provide further insight into how uncertainty-adjusted sorting operates jointly across firm characteristics and macroeconomic states. In particular, the interaction between firm size and aggregate volatility (SIZE × VXO) is positive and statistically significant, indicating that the effect of firm size on ranking adjustments is amplified during periods of elevated market uncertainty.

This result is economically intuitive. Larger firms typically feature richer information environments and more stable fundamentals, which makes their prediction uncertainty more structured rather than purely idiosyncratic. When aggregate uncertainty rises, the uncertainty-adjusted procedure becomes more conservative overall, compressing extreme ranks. Within this environment, large firms are better able to retain or improve their relative ranking under uncertainty-adjusted bounds because their predictions remain comparatively reliable. As a result, uncertainty-adjusted sorting reshapes rankings more strongly among large firms precisely when market-wide uncertainty is high.

Similar interaction patterns appear for profitability and investment characteristics. Positive interaction coefficients imply that uncertainty-adjusted sorting is not a mechanical volatility filter, but a selective mechanism that conditions ranking adjustments on the joint configuration of firm fundamentals and macroeconomic uncertainty. In high-uncertainty states, firms with characteristics associated with more reliable prediction signals experience larger relative ranking changes, while firms with less stable signals are more aggressively compressed toward the center of the ranking distribution.

Overall, the interaction effects demonstrate that uncertainty-adjusted sorting responds endogenously to macroeconomic conditions by reallocating ranking weight toward firms whose predictive signals are more robust when estimation risk is elevated.

\subsection{Placebo Test: Permuting Asset-Level Predictive Uncertainty}

\subsubsection{Motivation.}

The uncertainty-adjusted portfolio construction proposed in this paper relies on asset-specific predictive uncertainty, measured by residual-quantile half-widths derived from each stock's own historical prediction errors. By incorporating these uncertainty measures into the cross-sectional sorting rule, the strategy is designed to discipline extreme portfolio selections when predictions are empirically unreliable.

A central identification concern is whether the observed improvements in uncertainty-adjusted performance truly arise from the correct matching of uncertainty to individual assets, or whether they reflect more mechanical effects. Such alternative explanations include aggregate volatility states, time-series variation in overall prediction dispersion, or implicit regularization induced by perturbing the sorting rule, none of which require asset-level uncertainty information.

To disentangle these channels, we conduct a sequence of permutation (placebo) tests that construct counterfactual environments in which specific dimensions of uncertainty are selectively disabled while other features of the data remain unchanged. Performance differences across these tests therefore provide direct evidence on whether asset-specific predictive uncertainty is economically meaningful for portfolio construction.

\subsubsection{Test Design.}

We implement three complementary permutation tests, each designed to isolate a distinct dimension of predictive uncertainty.

\paragraph{Test 1: Cross-Sectional Permutation (Asset-Level Misalignment).} In the first test, for each month $t$, we randomly permute the residual-quantile half-widths across stocks in the cross-section, while keeping point predictions unchanged. As a result, the uncertainty measure originally associated with stock $i$ is reassigned to a randomly selected stock $j$ in the same month.

This design holds fixed the aggregate level and cross-sectional dispersion of predictive uncertainty in each period, as well as all time-series variation in uncertainty. However, it deliberately removes the asset-specific alignment between uncertainty and expected returns, which is the core channel through which uncertainty-adjusted sorting is hypothesized to operate.

If the Sharpe ratio gains documented earlier are driven primarily by market-wide volatility states or common uncertainty shocks, performance should remain largely intact under this permutation. Conversely, a deterioration in performance would indicate that correctly matching uncertainty to individual assets is essential.

\paragraph{Test 2: Time-Series Permutation Within Asset (Temporal Misalignment).} The second test preserves asset specificity but disrupts temporal alignment. For each stock, we randomly permute its residual-quantile half-widths across time, reassigning uncertainty estimates from one year to another for the same asset.

This procedure preserves each asset's unconditional distribution of predictive uncertainty, but severs the link between uncertainty and the contemporaneous information set or macroeconomic state. In other words, uncertainty remains stock-specific but is no longer synchronized with the conditions under which predictions are made.

This test isolates whether uncertainty-adjusted sorting benefits from time-varying, state-dependent uncertainty information, as opposed to relying solely on persistent differences in uncertainty across assets.

\paragraph{Test 3: Full Permutation (Asset and Time Misalignment).} In the third test, residual-quantile half-widths are fully permuted across both assets and time. An uncertainty estimate from any stock-year pair may be reassigned to any other stock-year pair.

This benchmark eliminates all systematic alignment between predictive uncertainty, asset identity, and economic states, leaving only purely mechanical perturbations of the sorting rule. Any residual performance under this test therefore reflects non-informational effects, such as stochastic smoothing or incidental regularization, rather than economically meaningful uncertainty information.

Together, the three permutation tests form a structured identification strategy. The first test removes cross-sectional alignment, the second removes temporal alignment, and the third removes both simultaneously. By comparing portfolio performance across these counterfactual environments, we can assess whether the gains from uncertainty-adjusted portfolio construction arise from asset-specific, state-dependent predictive uncertainty, rather than from aggregate volatility timing or mechanical features of the ranking procedure.

In the next section, we report the empirical results of these tests and show that the performance gains largely vanish once the economically meaningful alignment between uncertainty and assets is disrupted.

\subsubsection{Empirical Results.}

\afterpage{
\clearpage
\begin{landscape}
\begin{table}[ht]
    \centering
    \caption{Placebo Permutation Tests for Uncertainty-Adjusted Portfolio Sorting}
    \label{tab:placebo_test}
    \begin{threeparttable}
    \includegraphics[width=1.3\textheight]{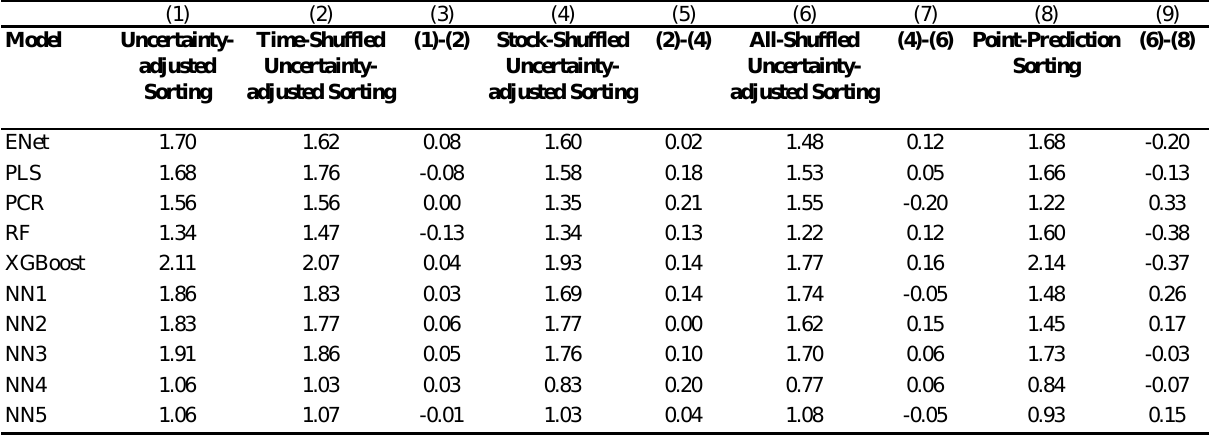}
        \begin{tablenotes}
            \item \scriptsize{\emph{Note.}This table reports annualized Sharpe ratios of monthly long–short portfolios formed using alternative ranking schemes over the 1992–2016 out-of-sample period. All portfolios are constructed using the same machine-learning point predictions; the sorting signal differs in how predictive uncertainty is incorporated. We consider four uncertainty-adjusted sorting schemes. No Shuffle (Column 1) uses asset-specific uncertainty measures correctly aligned with each stock’s own prediction residuals. Time Shuffle (Column 2) randomly permutes each stock’s uncertainty measures across time while preserving the cross-sectional identity of assets. Stock Shuffle (Column 4) randomly permutes uncertainty measures across stocks within each month, preserving time-series variation but breaking asset-level alignment. All Shuffle (Column 6) fully permutes uncertainty measures across both assets and time. The Point Prediction Sharpe ratio is reported as a benchmark. Four columns report differences: (i) No Shuffle minus Time Shuffle (Column 3), (ii) Time Shuffle minus Stock Shuffle (Column 5), (iii) Stock Shuffle minus All Shuffle (Column 7), and (iv) All Shuffle minus Point Prediction (Column 8). These differences isolate the incremental contribution of asset-level and time-series alignment of predictive uncertainty to uncertainty-adjusted portfolio performance.}
        \end{tablenotes}
    \end{threeparttable}

\end{table}
\end{landscape}
}

Table~\ref{tab:placebo_test} reports the results of permutation tests designed to isolate the economic role of asset-specific predictive uncertainty in uncertainty-adjusted portfolio construction. By systematically breaking the alignment between uncertainty estimates, assets, and time, these tests provide direct identification of the channels through which uncertainty adjustment affects risk-adjusted performance.

We have three observations. First, uncertainty-adjusted sorting with correctly aligned uncertainty estimates generally delivers the strongest Sharpe ratios across models. For most ML methods, the Sharpe ratio under the no-shuffle specification exceeds those obtained under shuffled alternatives. The ``No Shuffle -- Time Shuffle'' differences are typically positive, indicating that preserving the correct temporal alignment between uncertainty estimates and the information set contributes meaningfully to portfolio performance.

Second, disrupting temporal alignment weakens—but does not eliminate—the gains from uncertainty-adjusted sorting. When residual-based uncertainty measures are shuffled across time within each asset, Sharpe ratios decline relative to the no-shuffle benchmark but remain systematically higher than those obtained under cross-sectional or full shuffling. This pattern suggests that predictive uncertainty is time-varying and state dependent: uncertainty-adjusted sorting is most effective when uncertainty estimates are synchronized with prevailing market and macroeconomic conditions.

Third, breaking asset-level alignment eliminates most of the gains from uncertainty adjustment. Once uncertainty estimates are reassigned across stocks—either with or without additional time shuffling—performance deteriorates sharply. Sharpe ratios under the stock-shuffle and full-shuffle specifications are very similar and, for many models, approach those obtained under point-prediction sorting. Consistent with this, the ``Time Shuffle -- Stock Shuffle'' differences are generally positive, while the ``Stock Shuffle -- All Shuffle'' differences are small, indicating that asset-specific alignment is the dominant source of uncertainty-adjusted performance gains.

Finally, the mixed signs of the ``All Shuffle -- Point Prediction'' differences—positive for some models and negative for others—confirm that any residual performance under full shuffling reflects, at most, minor mechanical regularization effects rather than economically meaningful uncertainty information.

\subsubsection{Economic Interpretation.}

The permutation tests in Table~\ref{tab:placebo_test} provide clean identification evidence on the economic role of predictive uncertainty in uncertainty-adjusted sorting. The central insight is that the performance gains of uncertainty-adjusted portfolios arise primarily from correctly matching uncertainty estimates to individual assets; once this asset-specific link is disrupted, the strategy largely loses its advantage.

Across models, Sharpe ratios exhibit a clear and systematic \emph{pattern on average} across the four specifications. Performance is highest when uncertainty estimates are correctly aligned at both the asset and time dimensions (No Shuffle), declines when temporal alignment is disrupted but asset-level matching is preserved (Time Shuffle), deteriorates sharply when asset-level alignment is broken (Stock Shuffle), and becomes close to—or often below—point-prediction sorting under full shuffling (All Shuffle). While this ordering is not universal for every model comparison, it is highly consistent in the aggregate and provides clear identification of the mechanism through which uncertainty-adjusted sorting improves risk-adjusted performance.

The results show that \emph{asset-specific predictive uncertainty is essential}. When uncertainty estimates are correctly matched to individual stocks, uncertainty-adjusted sorting systematically improves Sharpe ratios relative to shuffled alternatives. Once uncertainty is reassigned across assets, performance collapses toward that of point-prediction sorting, indicating that the economic value of uncertainty adjustment is not driven by generic regularization or volatility timing but by asset-level information about estimation reliability.

Temporal alignment plays an additional but secondary role. When residual-based uncertainty measures are shuffled across time within each asset, Sharpe ratios decline relative to the no-shuffle benchmark but remain meaningfully higher than those obtained under cross-sectional or full shuffling. This intermediate performance indicates that predictive uncertainty is time-varying and state dependent: uncertainty-adjusted sorting is most effective when uncertainty estimates are synchronized with prevailing market and macroeconomic conditions.

By contrast, the near equivalence between stock-shuffle and full-shuffle outcomes demonstrates that cross-sectional alignment dominates temporal alignment. Once uncertainty is assigned to the wrong assets, preserving time-series variation alone contributes little incremental value. Finally, the mixed signs of the ``All Shuffle -- Point Prediction'' differences confirm that any residual performance under full shuffling reflects at most minor mechanical regularization effects rather than economically meaningful uncertainty information.

Taken together, these findings rule out explanations based solely on aggregate volatility timing, generic smoothing of rankings, or mechanical shrinkage. Instead, they confirm that uncertainty-adjusted sorting improves portfolio performance by incorporating economically meaningful, asset-level information about estimation reliability, thereby stabilizing extreme portfolio selections and reducing ranking noise.

\subsection{Uncertainty-Adjusted Sorting with Partial Uncertainty Information}

\begin{table}[ht]
    \TABLE{Uncertainty-Adjusted Bounds under Parametric Normal Approximation \label{tab:normal-interval-test}}
    {
    \begin{tabular}{@{}lcccc@{}}
\toprule
\textbf{Model} & \textbf{Point-Prediction Sorting} & \makecell{\textbf{1\% Normal-Approx.} \\ \textbf{Bound}} & \makecell{\textbf{5\% Normal-Approx.} \\ \textbf{Bound}}  & \makecell{\textbf{10\% Normal-Approx.} \\ \textbf{Bound}}  \\
\midrule
ENet & 1.68	& 1.69 & 1.72 & 1.69 \\
PLS & 1.66 &	1.66	& 1.75	& 1.85 \\
PCR & 1.22	& 1.29 &	1.59 &	1.56 \\
RF & 1.60 &	1.62	& 1.64	& 1.55 \\
XGBoost & 2.14	& 2.17 &	2.08 &	2.07 \\
NN1 & 1.48	& 1.85 &	1.91	& 1.92 \\
NN2 & 1.45	& 1.84 &	1.83 &	1.71 \\
NN3 & 1.73	& 1.72 & 	1.72	& 1.62 \\
NN4 & 0.84	& 1.15 &	1.20 &	1.11 \\
NN5 & 0.93	& 1.17	& 1.13 &	1.01 \\
\bottomrule
\end{tabular}
    }
    {\emph{Note.} This table reports annualized Sharpe ratios of long–short portfolios constructed using uncertainty-adjusted bounds derived from a parametric normal approximation. For each model, prediction residuals are first computed from rolling-window point predictions, and their standard deviation is estimated using historical calibration data. Symmetric upper and lower bounds are then constructed assuming conditional normality, using standard normal critical values corresponding to alternative two-sided coverage levels. Specifically, bounds are given by $\widehat{\mu}_{i,t} \pm z_{(1+\alpha)/2} \cdot \widehat{\sigma}_{i,t}$,  where $z_{(1+\alpha)/2}$ denotes the standard normal quantile corresponding to the nominal coverage level. Portfolios are formed each month by ranking stocks on the resulting uncertainty-adjusted upper bounds for the long leg and lower bounds for the short leg, following the same portfolio construction procedure as in the main analysis. The table compares the Sharpe ratios of these uncertainty-adjusted portfolios with those obtained from portfolios formed using point-prediction ranking alone.}
\end{table}

Table~\ref{tab:normal-interval-test} examines whether the performance gains from uncertainty-adjusted portfolio construction require detailed knowledge of the full empirical distribution of prediction residuals, or whether coarser uncertainty information is sufficient.

In the baseline analysis (Table~\ref{tab:ci_results}), uncertainty-adjusted bounds are constructed using asset-specific residual quantiles estimated nonparametrically from historical prediction errors. While this approach fully exploits the empirical distribution of residuals, it may be viewed as information intensive. Table~\ref{tab:normal-interval-test} therefore considers a deliberately simplified alternative in which uncertainty is summarized solely by the second moment of residuals, combined with a parametric normal approximation.

Formally, let $\widehat{\sigma}_{i,t}$ denote the sample standard deviation of the residual $\widehat{\varepsilon}_{i,t} = r_{i,t+1} - \widehat{\mu}_{i,t}$.
Assuming a symmetric normal approximation, we construct the uncertainty-adjusted bounds as
\[
\text{PI}_{i,t}^{(\text{N})}(\alpha) = \big[\, \widehat{\mu}_{i,t} - z_{(1+\alpha)/2} \cdot \widehat{\sigma}_{i,t}, \;
\widehat{\mu}_{i,t} + z_{(1+\alpha)/2} \cdot \widehat{\sigma}_{i,\tau}
 \, \big],
\]
where $z_{(1+\alpha)/2}$ denotes the $(1+\alpha)/2$ quantile of the standard normal distribution (e.g., $z_{0.505} = 0.0125$ for $\alpha = 1\%$, $z_{0.525} = 0.0627$ for $\alpha = 5\%$ and $z_{0.55} = 0.1257$ for $\alpha = 10\%$).

Using these parametric uncertainty-adjusted bounds, we form long–short portfolios by ranking stocks on the upper bound for the long leg and on the lower bound for the short leg, exactly as in the uncertainty-adjusted strategy described in Section~\ref{sec:port-constr}. Portfolio Sharpe ratios are reported for nominal coverage levels of 1\%, 5\% and 10\%, alongside the benchmark Sharpe ratios obtained from point-prediction-based sorting.

We have two observations. First, across eight out of ten models except XGboost and NN3, portfolios constructed using parametric normal distribution quantiles outperform point-prediction portfolios in terms of Sharpe ratio, despite relying on substantially less information about the residual distribution. Second, performance improvements are not monotone in the width of the bound: there exists a quantile level, which corresponds to larger values of $z_{(1+\alpha)/2}$, that tends to produce greater volatility reduction and higher uncertainty-adjusted returns compared to higher and lower quantile levels.

These results demonstrate that the economic value of uncertainty-adjusted portfolio construction does not hinge on precise estimation of the entire residual distribution. Even partial uncertainty information, summarized by a single scale parameter and combined with a generic parametric approximation, is sufficient to meaningfully improve ranking stability and portfolio efficiency. This finding reinforces the central mechanism of the paper: incorporating predictive uncertainty into the sorting rule enhances uncertainty-adjusted performance primarily by disciplining noisy extreme selections, rather than by exploiting fine-grained distributional features.

\section{Conclusion}\label{sec:conc}

This paper develops and empirically evaluates an uncertainty-adjusted-sorted asset pricing framework that explicitly incorporates predictive uncertainty into portfolio construction. Departing from the conventional focus on point predictions, we show that information contained in uncertainty-adjusted bounds---constructed from out-of-fold calibration errors---can be systematically exploited to improve uncertainty-adjusted portfolio performance.

Methodologically, we introduce a disciplined procedure for constructing asset-specific uncertainty-adjusted bounds under a rolling-window, time-respecting cross-validation scheme. The approach produces calibrated measures of predictive dispersion that are both model- and stock-specific, and that are insulated from in-sample overfitting. Importantly, our framework is model-agnostic: it applies uniformly to linear regularized regressions, dimension-reduction methods, tree-based algorithms, and neural networks. This generality allows us to study how predictive uncertainty interacts with model flexibility in a unified setting.

Empirically, we document several robust findings. First, while uncertainty-adjusted portfolio construction typically reduces raw return spreads relative to point-prediction sorting, it substantially lowers portfolio volatility. As a result, uncertainty-adjusted performance---as measured by Sharpe ratios and statistical significance---improves for the majority of models. These gains are particularly pronounced for flexible ML methods, such as neural networks, that are most susceptible to estimation noise under point-based ranking. Second, the improvements persist after accounting for transaction costs and in both long–short and benchmark-relative long-only settings, indicating that the results are economically meaningful rather than artifacts of leverage or short-selling.

We further provide direct evidence on the economic mechanism underlying these improvements. Firm-level and macroeconomic analyses show that uncertainty-adjusted sorting induces systematic and state-dependent reordering of assets, with larger effects for firms and periods characterized by elevated predictive uncertainty. Placebo and permutation tests demonstrate that the gains are not driven by aggregate volatility timing, common uncertainty shocks, or mechanical regularization effects. Instead, they arise from correctly matching uncertainty estimates to individual assets.

Finally, we show that the value of uncertainty information does not hinge on precise knowledge of the full residual distribution. Even coarse approximations---such as uncertainty-adjusted bounds constructed under a normality assumption using only second-moment information---yield materially higher Sharpe ratios than point-prediction portfolios. This finding underscores the robustness and practical relevance of the framework: incorporating even partial information about predictive uncertainty can meaningfully improve portfolio selection.

Taken together, our results suggest a simple but powerful lesson for empirical asset pricing and ML–based investment strategies. Predictive uncertainty is not merely a statistical nuisance to be averaged away; it is an economically meaningful object that should be incorporated directly into decision rules. uncertainty-adjusted portfolio construction offers a transparent and broadly applicable way to do so, improving ranking stability and uncertainty-adjusted performance without requiring stronger assumptions or more complex models. More broadly, the framework provides a bridge between modern ML and classical asset pricing concerns about estimation risk, offering a principled approach to uncertainty-aware investment decisions.

\bibliographystyle{informs2014} %
\bibliography{references} %

\ECSwitch

\ECHead{Additional Results}

To complement the analysis in Section~\ref{sec:rank-improvement},
we present panel fixed-effects regressions that relate ranking improvements of other ML models (PLS, PCR, RF, XGBoost, and NN1--NN5) to the firm-level and macro variables.

\afterpage{
\clearpage
\begin{landscape}
\begin{table}[ht]
    \centering
    \caption{Regression Evidence on Ranking Improvements from Interval Prediction: PLS}
    \label{tab:rank-PLS}
\includegraphics[width=1.35\textheight]{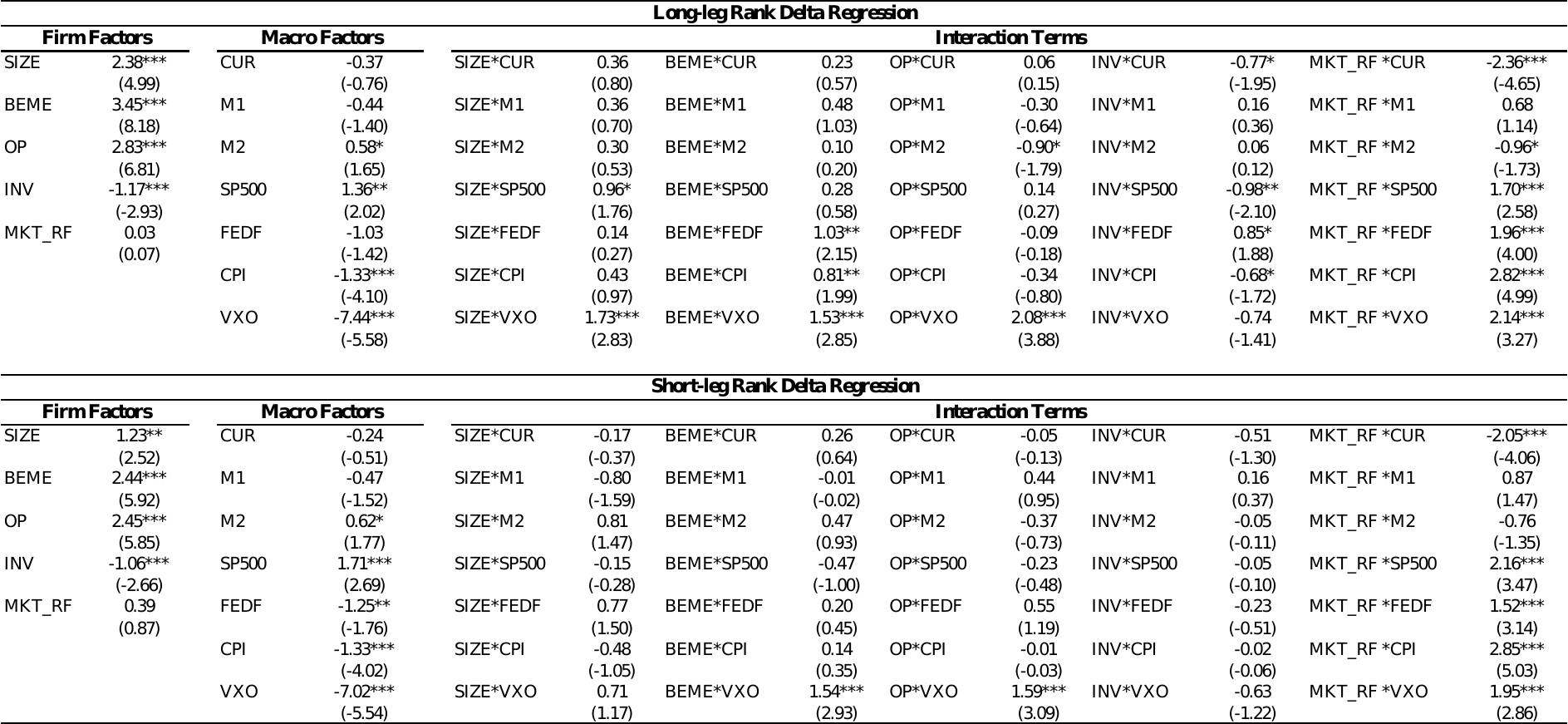}

\end{table}
\end{landscape}
}

\afterpage{
\clearpage
\begin{landscape}
\begin{table}[ht]
    \centering
    \caption{Regression Evidence on Ranking Improvements from Interval Prediction: PCR}
    \label{tab:rank-PCR}
\includegraphics[width=1.35\textheight]{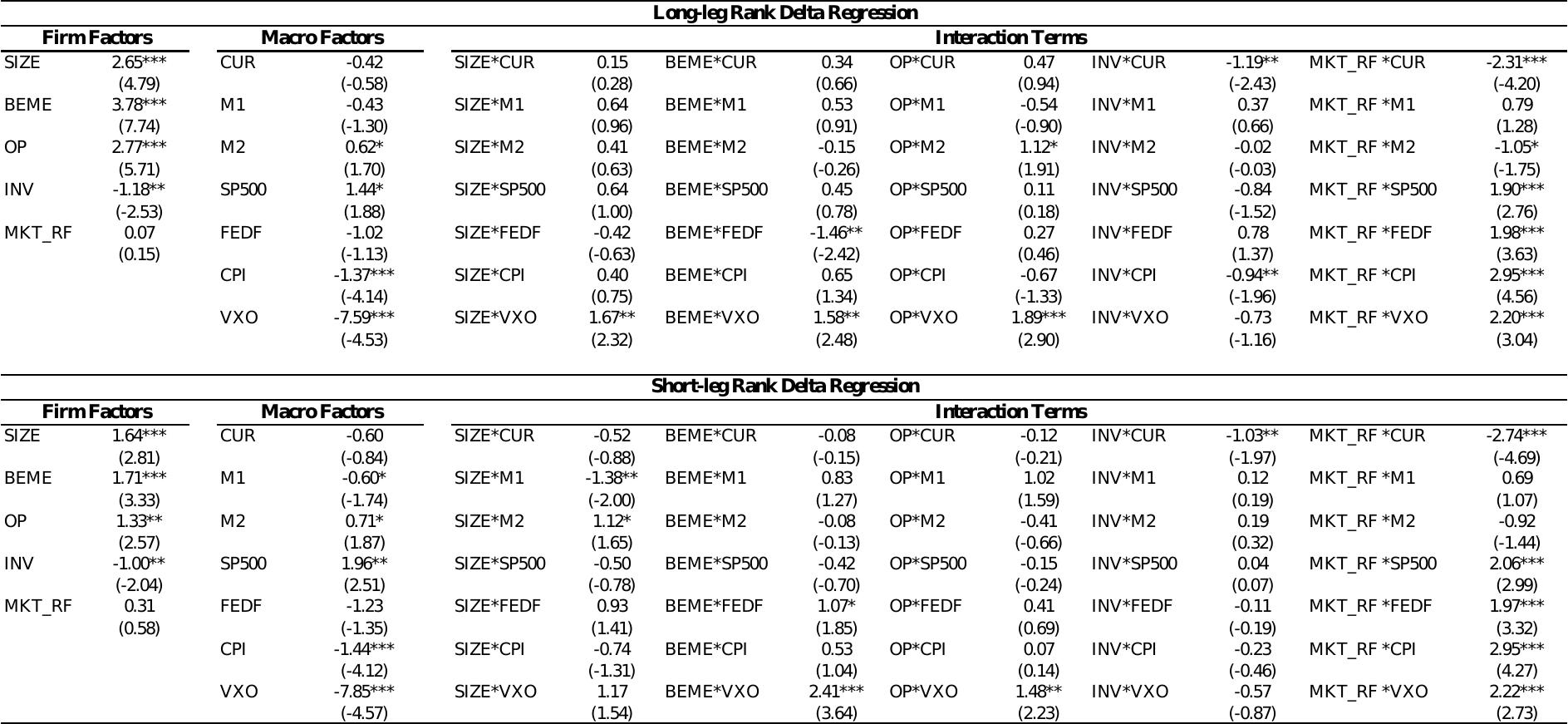}

\end{table}
\end{landscape}
}

\afterpage{
\clearpage
\begin{landscape}
\begin{table}[ht]
    \centering
    \caption{Regression Evidence on Ranking Improvements from Interval Prediction: RF}
    \label{tab:rank-RF}
\includegraphics[width=1.35\textheight]{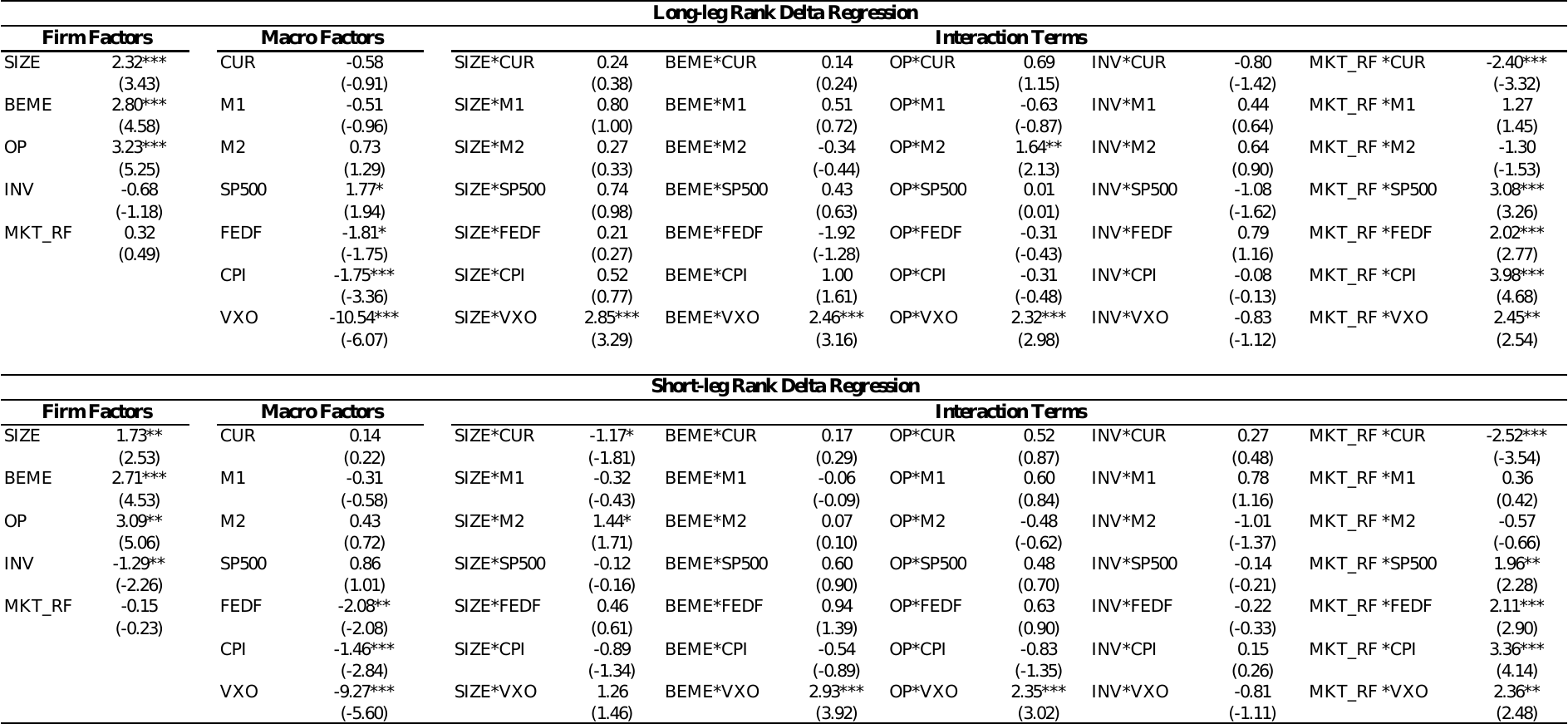}

\end{table}
\end{landscape}
}

\afterpage{
\clearpage
\begin{landscape}
\begin{table}[ht]
    \centering
    \caption{Regression Evidence on Ranking Improvements from Interval Prediction: XGBoost}
    \label{tab:rank-XGB}
\includegraphics[width=1.35\textheight]{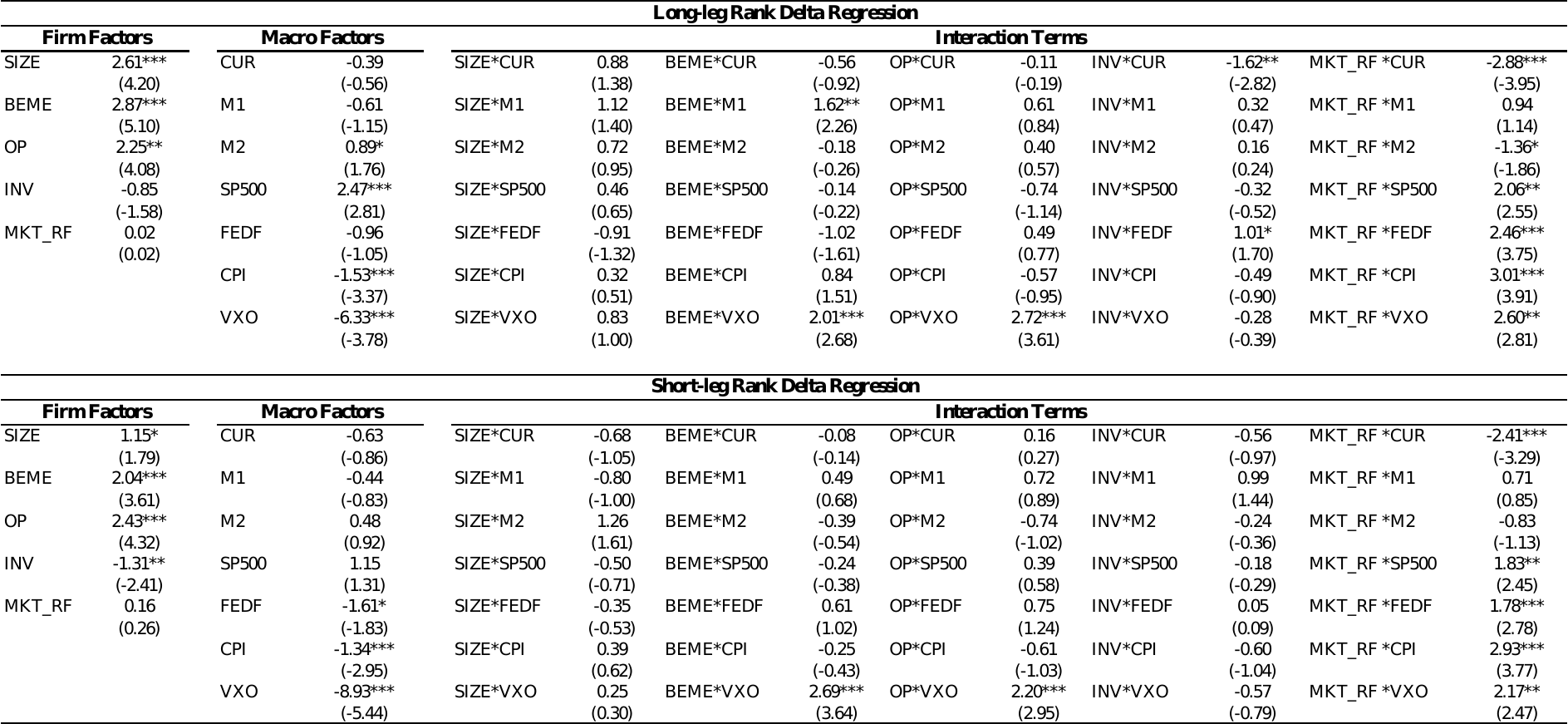}

\end{table}
\end{landscape}
}

\afterpage{
\clearpage
\begin{landscape}
\begin{table}[ht]
    \centering
    \caption{Regression Evidence on Ranking Improvements from Interval Prediction: NN1}
    \label{tab:rank-NN1}
\includegraphics[width=1.35\textheight]{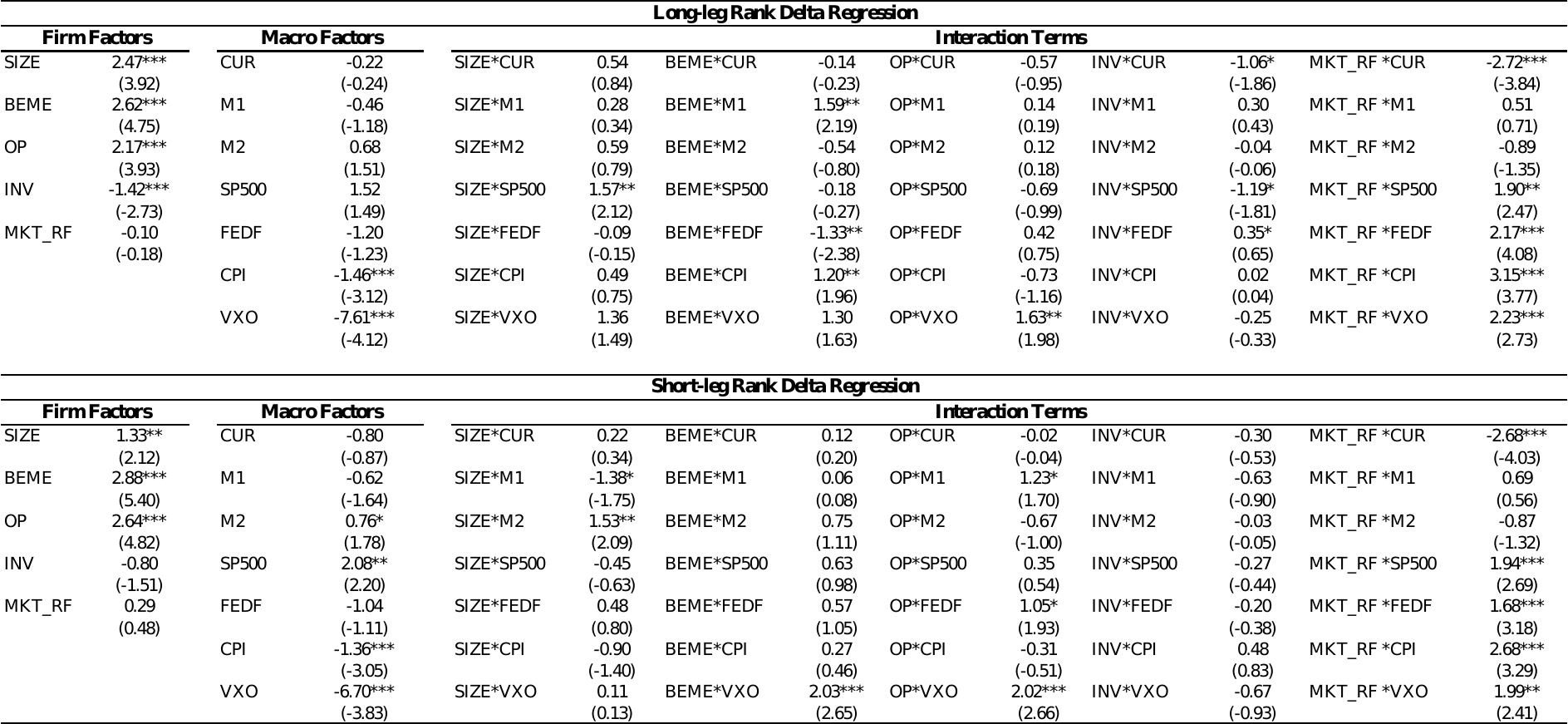}

\end{table}
\end{landscape}
}

\afterpage{
\clearpage
\begin{landscape}
\begin{table}[ht]
    \centering
    \caption{Regression Evidence on Ranking Improvements from Interval Prediction: NN2}
    \label{tab:rank-NN2}
\includegraphics[width=1.35\textheight]{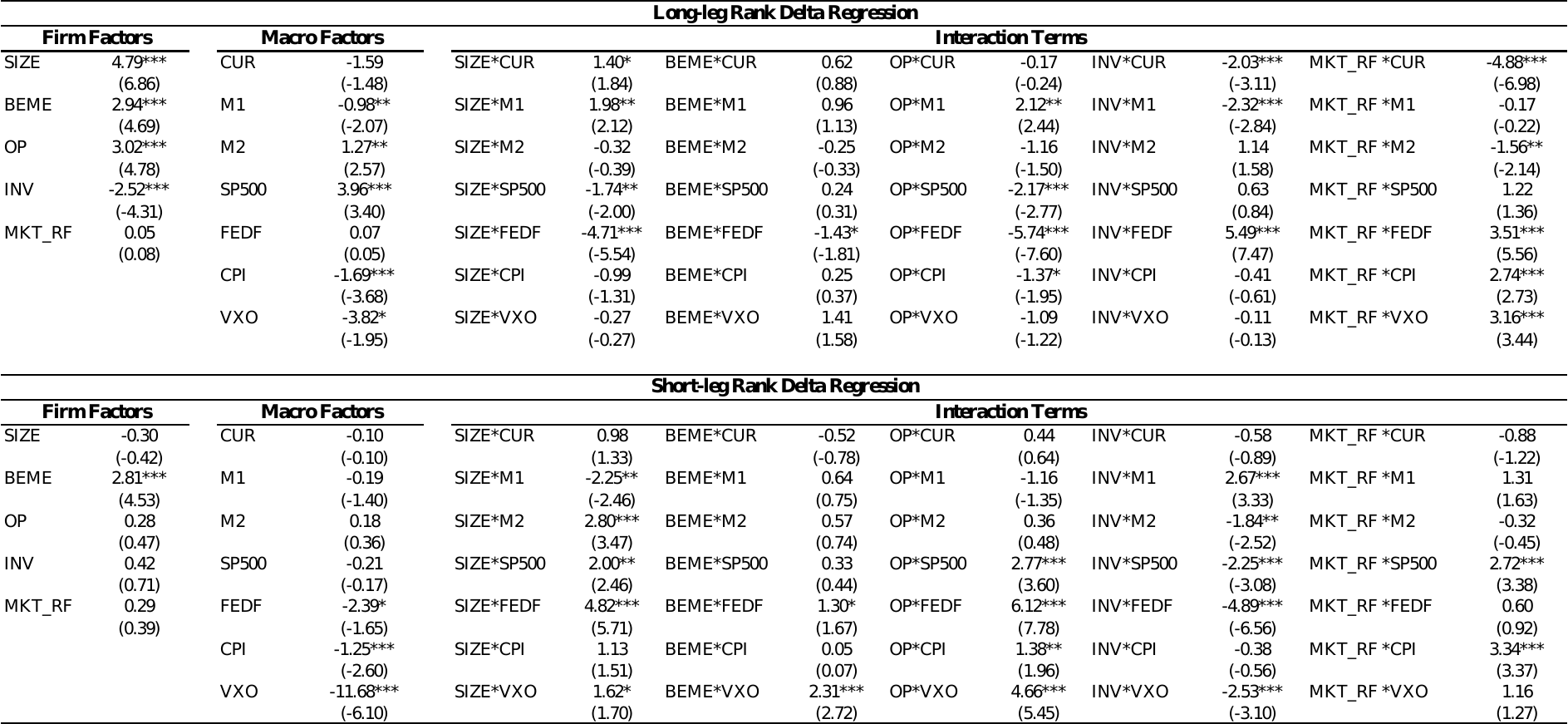}

\end{table}
\end{landscape}
}

\afterpage{
\clearpage
\begin{landscape}
\begin{table}[ht]
    \centering
    \caption{Regression Evidence on Ranking Improvements from Interval Prediction: NN3}
    \label{tab:rank-NN3}
\includegraphics[width=1.35\textheight]{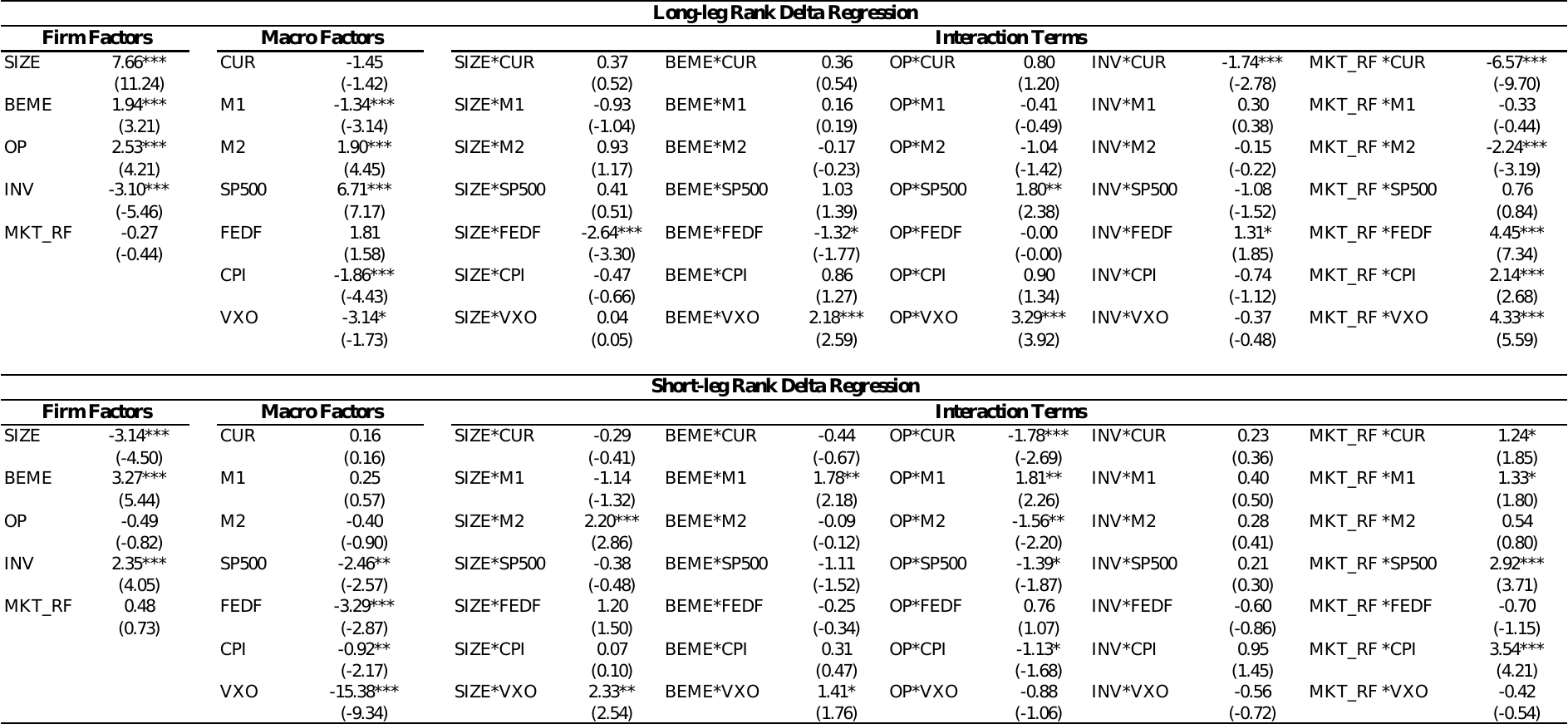}

\end{table}
\end{landscape}
}

\afterpage{
\clearpage
\begin{landscape}
\begin{table}[ht]
    \centering
    \caption{Regression Evidence on Ranking Improvements from Interval Prediction: NN4}
    \label{tab:rank-NN4}
\includegraphics[width=1.35\textheight]{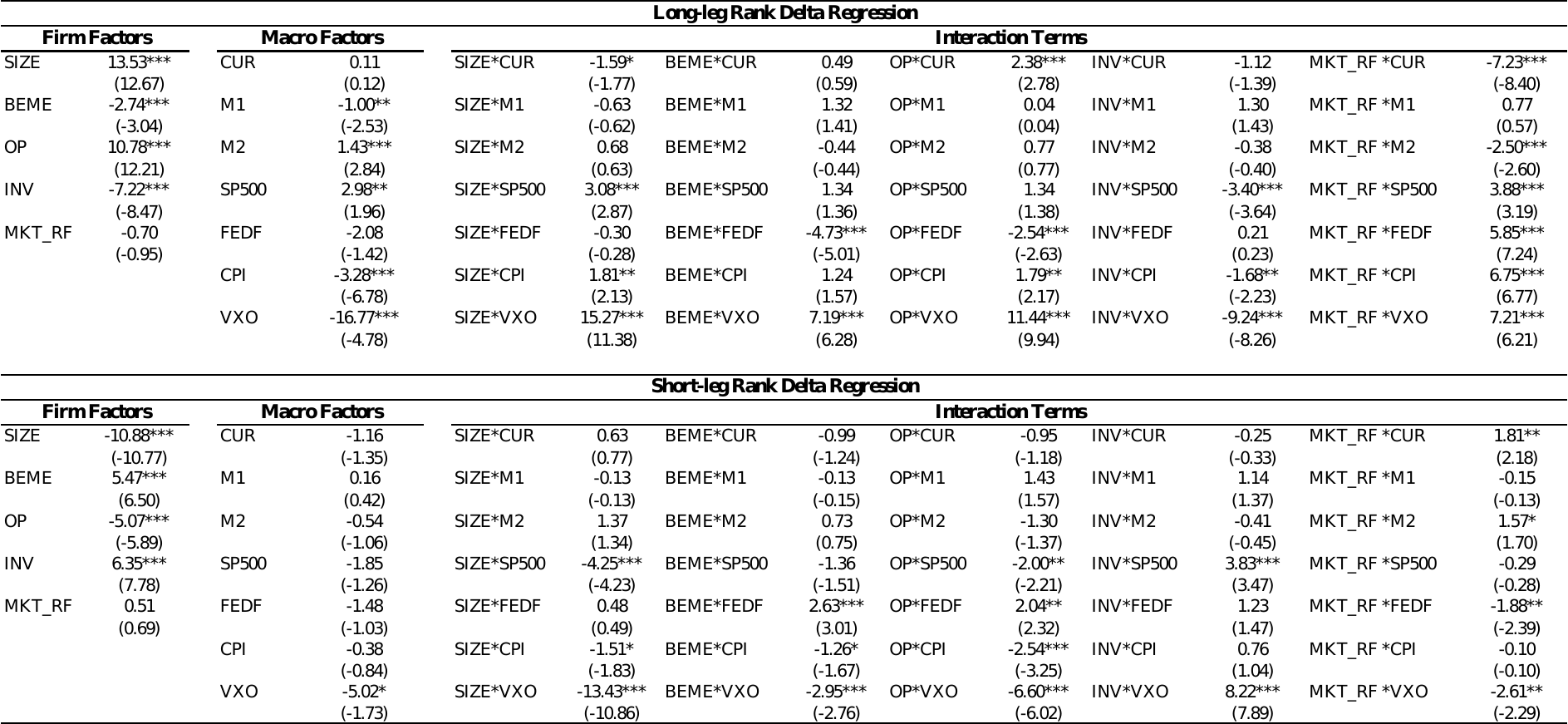}

\end{table}
\end{landscape}
}

\afterpage{
\clearpage
\begin{landscape}
\begin{table}[ht]
    \centering
    \caption{Regression Evidence on Ranking Improvements from Interval Prediction: NN5}
    \label{tab:rank-NN5}
\includegraphics[width=1.35\textheight]{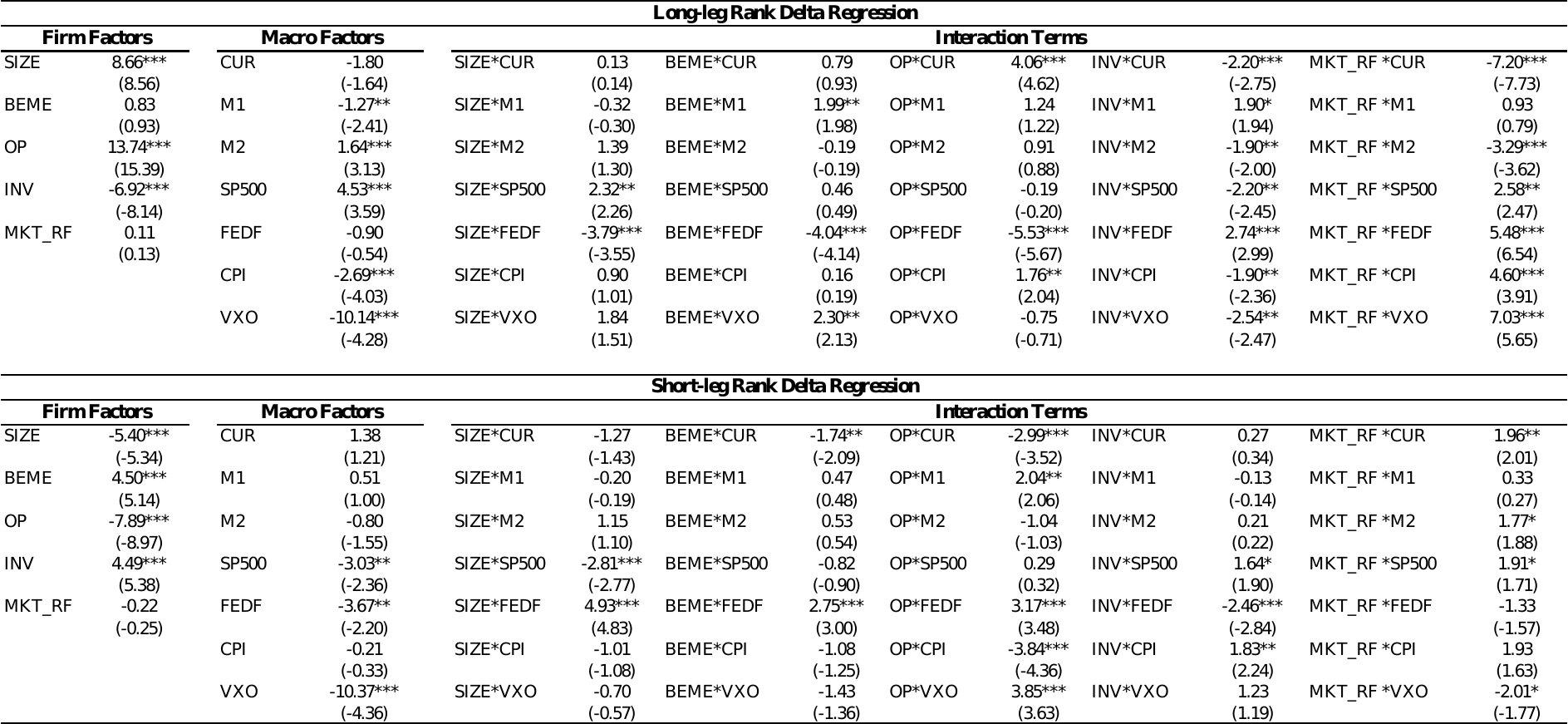}

\end{table}
\end{landscape}
}

\end{document}